\RequirePackage{tikz} 
\documentclass{jaa}

\usepackage{graphicx}

\usepackage{graphicx,amsmath}

\usepackage{caption}
\usepackage{subfigure}
\usepackage{xcolor}
\usepackage{graphicx}
\usepackage{natbib}
\usepackage{color}
\usepackage{pdflscape}
\usepackage{hyperref}
\usepackage{float}
\usepackage{balance}
\usepackage[latin1]{inputenc}
\usepackage{soul}
\setstcolor{blue}
\usepackage{tikz}
\usepackage{amssymb}
\usetikzlibrary{shapes,arrows,positioning}
\let\new=\newcommand 
\new{\diff}{{\rm d}}
\renewcommand\[{\begin{equation}}
\renewcommand\]{\end{equation}}

\def\myfnt{\ifx\protect\@typeset@protect\expandafter\footnote\else\expandafter\@gobble\fi}

\newcommand{\araa}{{\it Ann. Rev. Astron. Astrophys.}}

\newcommand{\aap}{{\it Astron. Astrophys.}}

\newcommand{\aj}{{\it Astron. J.}} 
\newcommand{\apj}{{\it Astrophys. J.}}
\newcommand{\apjl}{{\it Astrophys. J. Lett.}}
\newcommand{\apjs}{{\it Astrophys. J. Suppl.}}
 
\newcommand{\cjaa}{{\it Chin. J. Astron. Astrophys.}}

\newcommand{\mnras}{{\it Mon. Not. Roy. Astron. Soc.}}

\begin{document}

\title{Tracing the outer disk of NGC~300: An ultraviolet view}


\author{Chayan Mondal\textsuperscript{1,2,*}, Annapurni Subramaniam\textsuperscript{1}, Koshy George\textsuperscript{3}}
\affilOne{\textsuperscript{1}Indian Institute of Astrophysics, 2nd Block, Koramangala, Bangalore 560034\\}
\affilTwo{\textsuperscript{2}Pondicherry University, R.V. Nagar, Kalapet, 605014, Puducherry, India\\}
\affilThree{\textsuperscript{3}Department of Physics, Christ University, Bangalore, India}


\twocolumn[{
\maketitle

\corres{chayan@iiap.res.in}

\msinfo{12 March 2019}{11 July 2019}{}

\begin{abstract}
We present an ultra-violet (UV) study of the galaxy NGC~300 using GALEX far-UV (FUV) and near-UV (NUV) observations. We studied the nature of UV emission in the galaxy and correlated it with optical, H~I and mid-infrared (3.6 $\mu$m) wavelengths. Our study identified extended features in the outer disk, with the UV disk extending up to radius 12 kpc ($>$2R$_{25})$. We estimated the FUV and NUV disk scale-length as 3.05$\pm$0.27 kpc and 2.66$\pm$0.20 kpc respectively. The scale-length in FUV is 2.3 times larger than that at 3.6 $\mu$m, and we also find the disk to gradually become flatter from longer to shorter wavelengths. We performed a statistical source subtraction to eliminate the background contaminants and identified 261 unresolved UV sources between the radii 5.3 kpc and 10 kpc (1 $\sim$ 2 R$_{25}$). The identified UV sources show an age range between 1 - 300 Myr with a peak at 25 Myr and a mass range between $10^3 M_{\odot}$ to $10^6 M_{\odot}$, estimated using starburst99 models. The north-eastern spiral arm is found to be populated by young low mass sources suggesting that the star formation in this spiral arm is a recent phenomenon. The UV emission beyond the R$_{25}$ radius has contribution from these low mass sources and is extended up to $\sim$ 2R$_{25}$ radius. We conclude that NGC~300 has an extended UV disk, mainly populated by young low mass sources. The star formation rate is measured to be $\sim$0.46 $M_{\odot}/yr$ which is comparable to its near optical twin M33.

\end{abstract}
}]
\doinum{}
\artcitid{}
\volnum{}
\year{2019}
\pgrange{1--15}
\setcounter{page}{1}
\lp{15}

\section{Introduction}
Evolution of galaxies over a time scale is dictated by its star formation. The star formation in the outer disks
of galaxies has attracted a lot of attention recently, where these studies are centered around the nearby galaxies (Thilker et al. 2005a,b, 2007; Gil de Paz et al. 2005). 
Outer parts of galaxies where the gas densities are relatively low, present challenges to our understanding of the cloud/star formation processes. Also, deciphering the star formation histories of 
outer stellar disks provide information on the growth of galaxy disks with time and 
tests current models of disk growth (see, for example, Azzollini et al. (2008)). Understanding star formation in the outer stellar disk of galaxies is also important for chemical enrichment as well as the impact of stellar feedback into the low density interstellar medium (ISM) (Thilker et al. 2005b). Different proxies are used to trace star formation history in a galaxy. Massive young OB stars embedded in molecular clouds show $H_{\alpha}$ 
emission which traces very recent star formation up to $\sim$ 10 Myr (Knapen et al. 2006; Kennicutt 1998). Since tracing  CO emission, which is an
indicator of molecular hydrogen, is difficult in galaxies beyond the local group, $H_{\alpha}$ emission is in general compared
with the H~I column density maps. Both $H_{\alpha}$ and H~I are good tracers to
infer the on-going star formation and location of star forming material. Far-ultraviolet (FUV) radiation which traces the location of 
massive stars, is used as another
proxy to trace star formation in combination with $H_{\alpha}$ emission (Goddard et al. 2010). H$\alpha$ emission traces present day star formation, 
only up to a few Myr ago. On the other hand, ultraviolet (UV) flux and UV colours, 
used in this study, can trace star formation up to a few hundred Myr (Kennicutt \& Evans 2012).\\

The presence of extended UV disk (XUV disk) in 
many nearby galaxies was confirmed through UV observations by GALEX (Thilker et al. 2005a, 2007; Gil de Paz et al. 2005; Zaritsky \& Christlein 2007; Wilsey \&
Hunter 2010). 
It was first discovered in the spiral galaxy M83 by Thilker et al. (2005b). Soon after, a  few other
galaxies with XUV disks well beyond their classical optical radius ($R_{25}$, the radius at which surface brightness 
of the galaxy falls below 25 mag/arcsec$^2$ in B band), were also found. 
Star forming UV knots were identified even up to $4R_{25}$ radius in M83 (Goddard et al. 2010). XUV
disks are classified into two categories by Thilker et al. (2007); those with structured, filamentary UV emission with
spiral patterns were classified as XUV I disks and those having large UV emission in the outer disk were classified 
as XUV II disks. They identified NGC~300 to have a wispy extension of the inner disk at markedly low intensity.
Goddard et al. (2010) compared FUV,
near-ultraviolet (NUV) and $H_{\alpha}$ measurements for
star forming regions in 21 galaxies, in order to characterize the properties of their disks at radii
beyond the main optical radius ($R_{25}$). They measured $H_{\alpha}$ flux profile along with UV and noticed that for 10
galaxies though $H_{\alpha}$ emission was truncated after $R_{25}$, the UV emission was detected up to a larger radius,
which confirmed that these galaxies have extended outer disks.
Zaritsky \& Christlein (2007) examined 11 galaxies and found an excess of blue ((FUV$-$NUV) $<$ 1, NUV $<$ 25)
sources out to $2R_{25}$  for $\sim$ 25 \% of their sample. Barnes et al. (2011) performed an 
analysis of ultra-deep UV and $H_{\alpha}$ imaging of five nearby spiral galaxies to study the recent star
formation in the outer disk and found UV flux extending up to 1.2-1.4 $R_{25}$ for most of them. Goddard et al. (2010) used starburst99 simple stellar population (SSP) model generated NUV magnitudes and (FUV$-$NUV) colors to determine masses and ages of identified 
clusters in two XUV disk galaxies, NGC 3621 and M83.\\

The Sculptor group is among the nearest galaxy groups beyond the Local Group (Jerjen et al. 1998), which 
is at a distance of 2 to 5 Mpc. The nearest sub group consists of NGC 55, NGC~300, and possibly two or more known 
spheroidal companions (Karachentsev
et al. 2003). NGC~300,
the brightest galaxy in the sculptor group, is nearly isolated, only a dwarf galaxy is found nearby (Tully et al. 2006; Karachentsev et al.
2003). 
This galaxy is also a near-optical twin of the Local Group galaxy M33. Bland-Hawthorn et al. (2005)
reported a pure exponential stellar disk up to $\sim$ 14.4 kpc 
($\sim 2.2R_{25}$), which is about 10 disk scale length
for NGC~300 while M33 has a disk break at $\sim$8 kpc (Ferguson et al. 2007; Barker et al. 2011). Present star formation rate (SFR) in the disk of NGC~300 has been measured and a relatively low value of 
$\sim$ 0.08 - 0.30 $M_{\odot}$ yr$^{-1}$ is estimated by different tracers like H$\alpha$ emission 
(Helou et al. 2004; Karachentsev \& Kaisina 2013), FUV luminosity (Karachentsev \& Kaisina 2013), mid-IR (Helou et al. 2004) and
X-ray (Binder et al. 2012). The standard model of disk galaxy evolution suggests an inside-out growth of the disk, i.e.
the inner part of the disk forming earlier than the  outer part.
Using HST observation of individual stars up to 5.4 kpc, Gogarten et al.
(2010) concluded an inside-out disk growth
in NGC~300 in the last 10 Gyr. Vlaji{\'c} et al. (2009) found an extended stellar disk 
and an upturn in metallicity gradient at a radius of $\sim$10 kpc in NGC~300 and concluded that 
radial mixing or accretion in the 
outer disk may be the reason behind this change in the metallicity gradient. Gogarten et al. (2010) also found a
similar metallicity gradient in the inner disk of the galaxy
through HST observation and suggested that the probability of radial mixing is very less in NGC~300 because 
of its low mass. Using HST observation, Hillis et al. (2016) studied the star formation history in the past 200 Myr 
in four different regions of NGC~300 and identified an unbroken young stellar disk at least 
up to 8 disk scale length. In another study, Rodr{\'i}guez et al. (2016) identified 1147 young stellar groups by studying
six different regions in the galaxy and noticed that these groups are mostly present along the spiral arm of the galaxy.\\

H~I disk of this galaxy is well studied and found to show some noticeable structures. Puche et al. (1990) reported a warp 
in the H~I disk just outside the optical disk. Westmeier et al. (2011) 
used ATCA radio observation and mapped the H~I disk of NGC~300 outward to a larger extent. They 
found  a dense inner disk and an extended outer disk of 35 kpc in diameter along the major axis of the galaxy. They also 
observed an asymmetry in the outer disk, which they speculated to be as due to ram pressure caused by the 
interaction between the galaxy and surrounding inter-galactic medium (IGM).\\

In this study, we used deep GALEX observations of NGC~300 in both FUV and NUV pass bands to understand the recent star formation in the galaxy. We studied the extent of FUV emission of the galaxy and correlated the overall UV disk structure with optical, H~I and infra-red emissions. These correlations portray the nature of young star forming regions in the galaxy. We generated radial profile for surface luminosity density and SFR density to understand the nature of UV emission. We also identified several UV sources in the outer disk of the galaxy and estimated their mass and age with the help of synthetic UV magnitudes/colours and explored their spatial distribution.\\ 

The paper is arranged as follows. The data are presented in Section \ref{s_data}, background and foreground source subtraction is discussed in
Section \ref{s_background}, reddening and metallicity in Section \ref{s_reddening},
models are in Section \ref{s_ssp}, analysis in section \ref{s_analysis} followed by a discussion and summary in
Sections \ref{s_discussion} and \ref{s_summary} respectively.

\section{Data}
\label{s_data}

The archival GALEX images of NGC~300, with a total exposure time of 12987.6 sec (NUV and FUV), 
were obtained from the MAST data archive. 
GALEX has two channels far-UV (1350-1750) $\AA$ and near-UV (1750-2800) $\AA$ with a resolution of 4.5-6$^{\prime \prime}$. These observations were obtained between 2004-10-26 to 2004-12-15, 
and the Image ID is 3073040579138954826 (tile name GI1\_061002\_NGC0300).
We used images as well as the catalog of sources produced by GALEX data pipeline (Morrissey et al.
2007) for the whole 1.25$^\circ$ field of view. The catalog contains 19289 identified sources within the total field
of view. 
Out of these, we found 6554 sources to have both FUV and NUV detection.  The catalog provides apparent magnitudes of sources in both FUV and NUV band
along with their RA and DEC.\\

We have also used Australia Telescope Compact Array (ATCA) \& Galactic All-Sky Survey (GASS)
combined H~I data of NGC~300 for correlating with UV properties.
The data cube, containing 53 image files, was obtained from Westmeier et al. (2011). We combined all the images using IDL and 
obtained a complete H~I map of the galaxy. In order to compare our results with optical and infra-red, we have used Digitized Sky Survey (DSS) blue band, Infrared Array Camera (IRAC) 3.6 $\mu$m and Multiband Imaging Photometer (MIPS) 24 $\mu$m images of the galaxy obtained from the NASA/IPAC Extragalactic Database (NED).\\

\begin{table}
\centering
\caption{Properties of NGC~300}
 \label{ngc300}
\resizebox{90mm}{!}{

\begin{tabular}{ccc}
\hline
 Property & Value & Reference\\\hline
 RA & 00:54:53.4 & Skrutskie et al. (2006)\\
 DEC & -37:41:03.7 & Skrutskie et al. (2006)\\
 Morphological type & SA(s)d & de Vaucouleurs et al. (1991)\\
 Distance & 1.9 Mpc & Rizzi et al. (2006)\\
 Inclination & $42.3^\circ$ & Puche et al. (1990)\\
 PA of major axis & $109^\circ$ & de Vaucouleurs \& Page (1962)\\
 Mass & $2.9\times10^{10} M_{\odot}$ & Westmeier et al. (2011)\\
 $R_{25}$ & 5.3 kpc & Faesi et al. (2014)\\
 Optical scale-length ($R_{d}$) &  2.1 kpc &  Carignan et al. (1985)\\\hline

\end{tabular}
}
\end{table}

\section{Background and foreground sources}
\label{s_background}

\begin{figure}
\begin{center}
\includegraphics[width=3.2in]{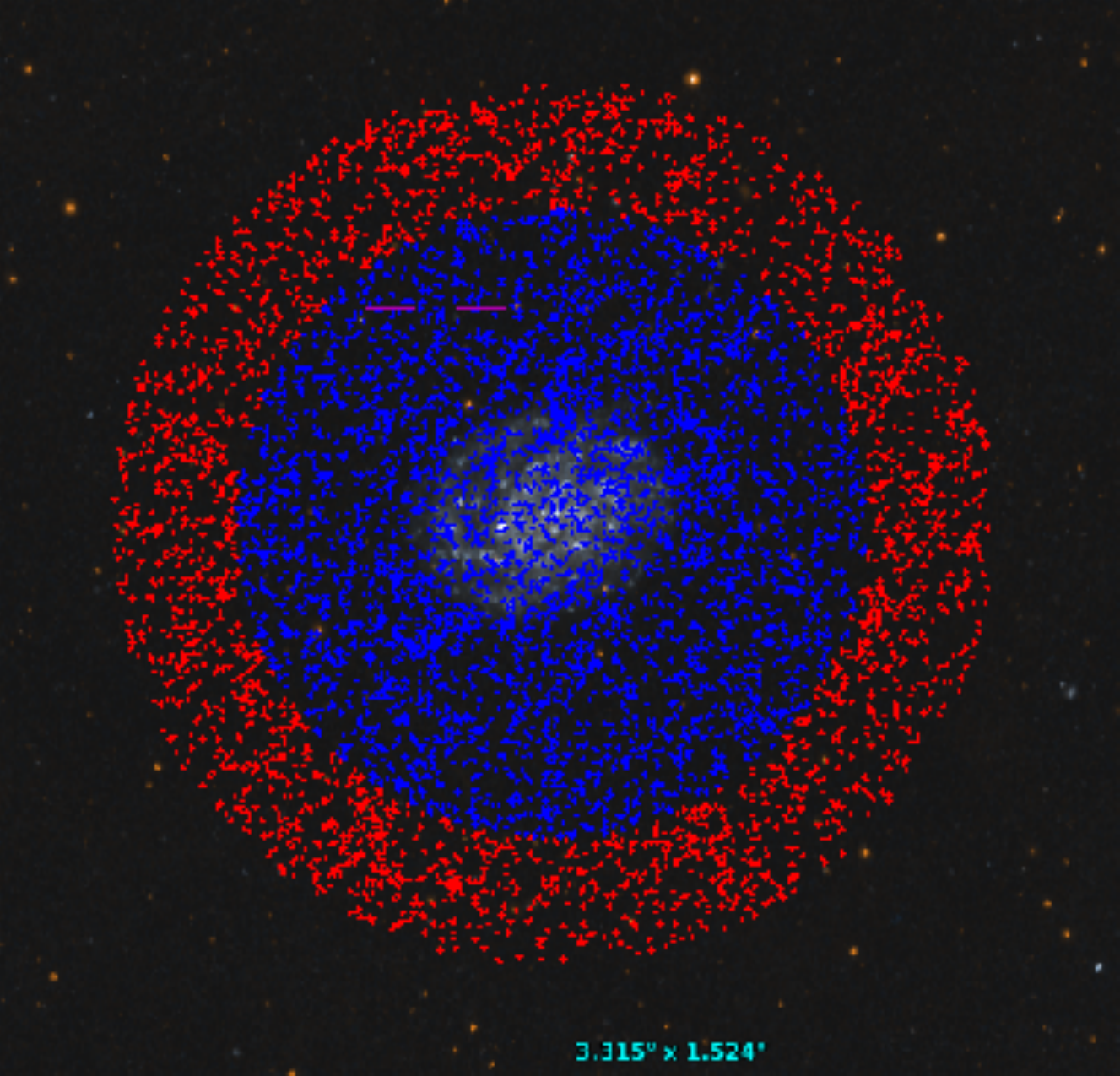} 
 \caption{The identified sources (detected in both FUV and NUV bands) are over plotted on GALEX image of NGC~300.
 The inner circular region, containing the blue points, is considered as the galaxy and outer annular region with 
 red points is assumed as field region.}
 \label{ngc300_in_out}
 \end{center}

\end{figure}

\begin{figure}
\centering
\subfigure[]{\includegraphics[width = 3.7in]{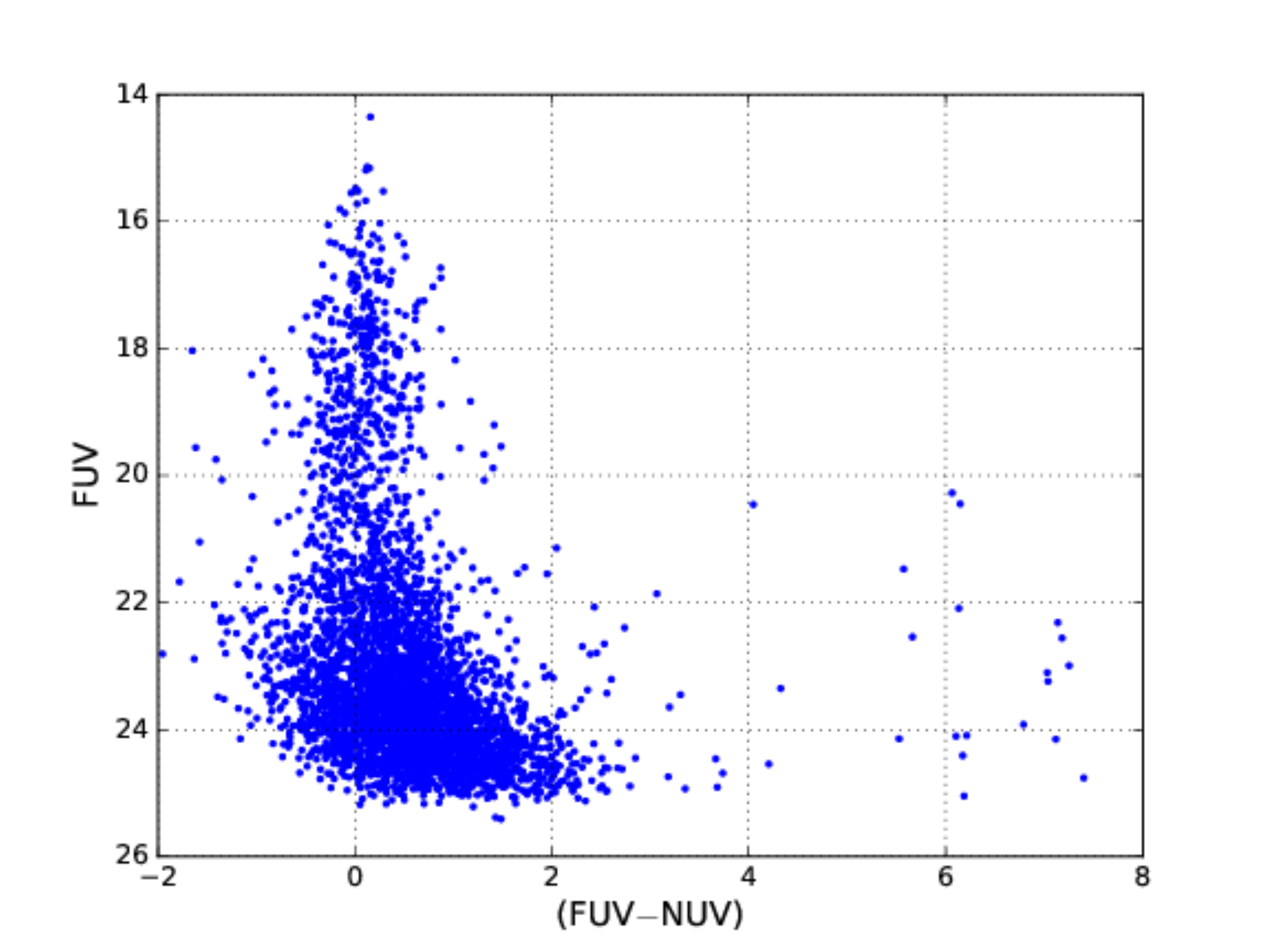}}
\subfigure[]{\includegraphics[width = 3.7in]{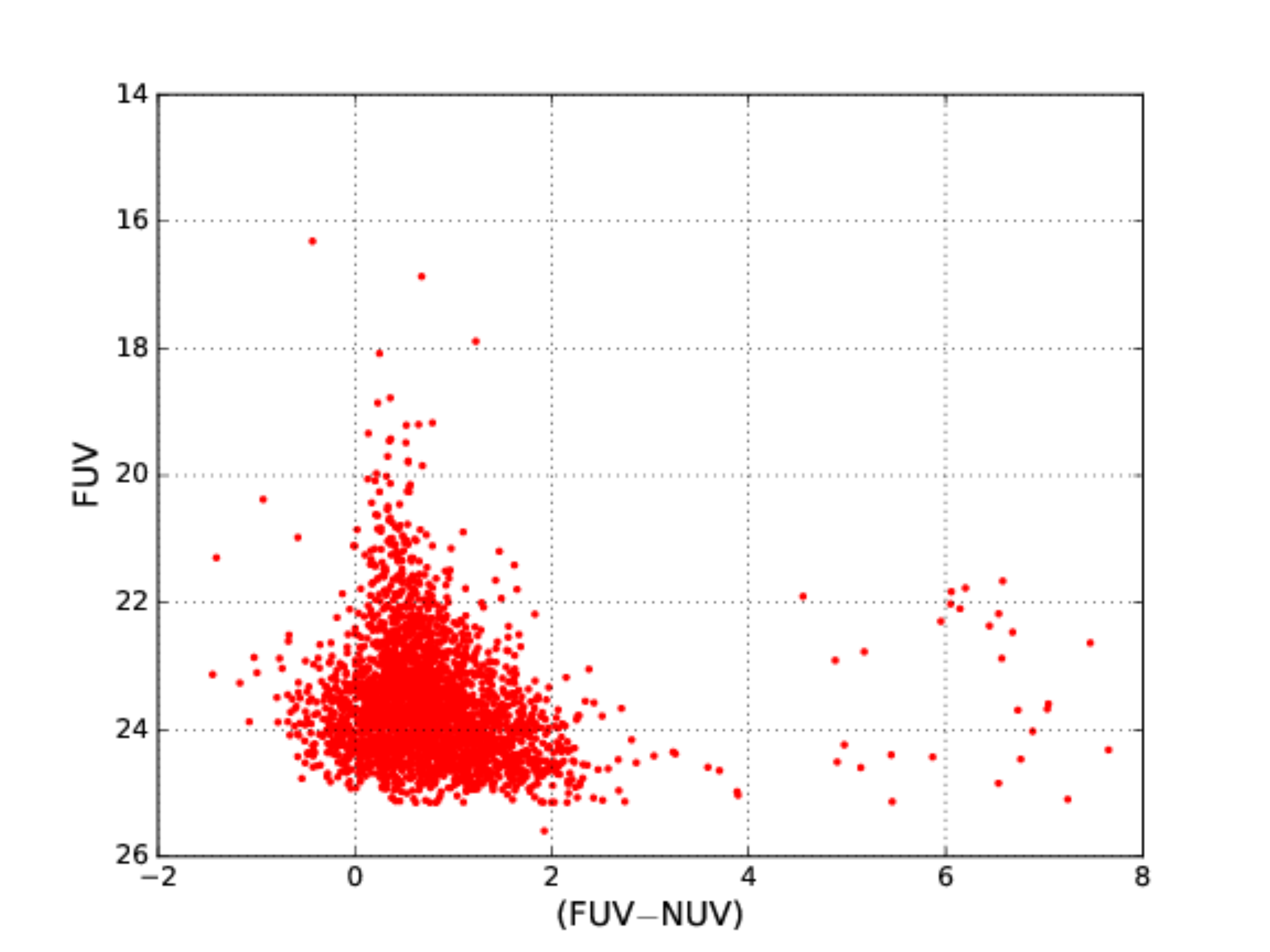}}
\caption{Figure (a) shows the FUV vs (FUV$-$NUV) CMD for galaxy region (blue points in Figure \ref{ngc300_in_out}) and Figure (b) shows
the CMD for outer field region (red points in Figure \ref{ngc300_in_out}).}
\label{cmd_in_out}
\end{figure}

\begin{figure}
\begin{center}
\subfigure[]{\includegraphics[width = 3.7in]{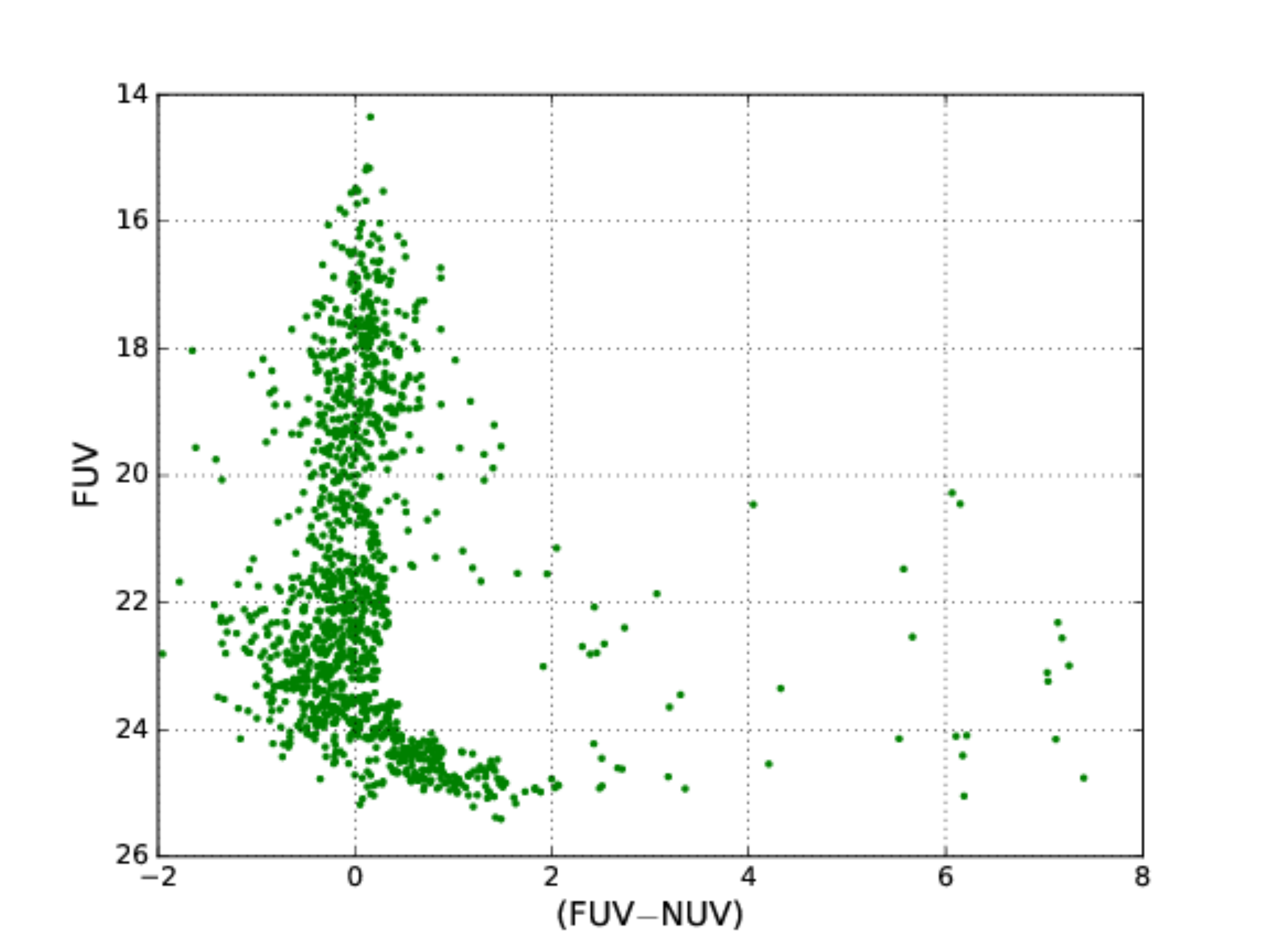}}
\subfigure[]{\includegraphics[width = 3.7in]{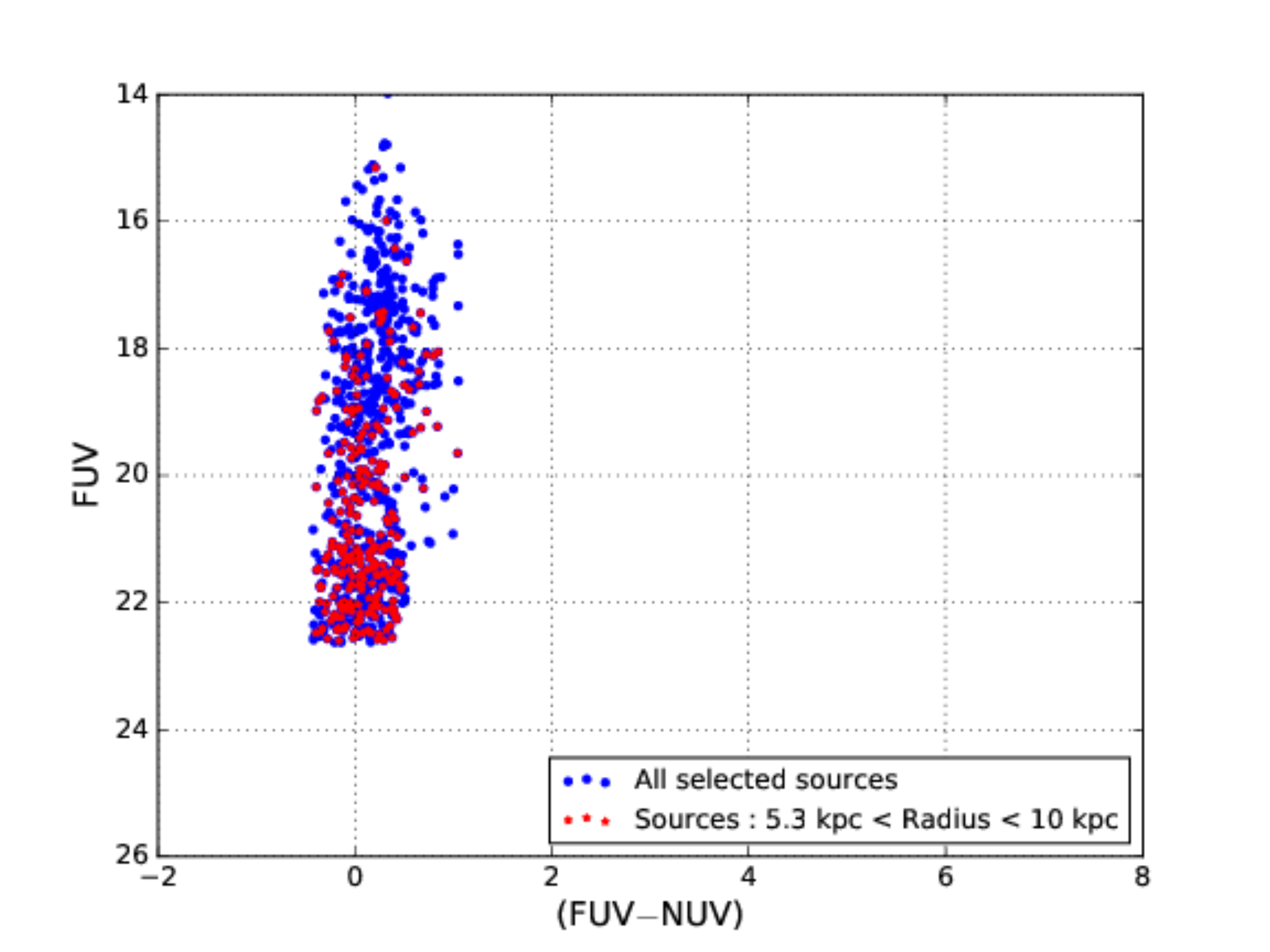}}
\caption{Figure (a) shows the FUV vs (FUV$-$NUV) CMD of sources remaining in the galaxy region after performing background
subtraction.
Figure (b) shows the CMD of 742 UV sources (reddening and extinction corrected) which
are finally considered as part of the galaxy. The sources marked with red star are present in the outer disk between radius 5.3 kpc to 10 kpc.}
\label{cmd_clean}
\end{center}
\end{figure}

It is very important to exclude background galaxies and foreground stars
from the identified source catalog.
Since the galaxy
NGC~300 is located at a high galactic latitude ($b\approx-79^\circ$) (Vlaji{\'c} et al. 2009), very less number of foreground stars
are expected to be present in the field. In order to remove background sources, we divided the whole GALEX tile into two regions having equal area which
are shown in Figure \ref{ngc300_in_out}. 
One is the inner circular region (i.e. galaxy region) having a radius of 25.8 arcmin and 
another is the outer annular region (i.e. field region) with an inner 
and outer radii of 25.8 and 36.5 arcmin
respectively. We assumed all the sources present in the outer annular region as background contaminants and used them 
to remove background sources present in the inner circular region by a statistical technique. We constructed
FUV vs (FUV$-$NUV) color magnitude diagrams (CMD) for sources present in both the galaxy region and the field region (Figure \ref{cmd_in_out}). 
We considered each source in the field CMD and identified its nearest counterpart in the galaxy CMD and then removed it from 
the inner circular region.
In order to do this, we constructed a grid of
[magnitude, color] bins with different sizes, starting from [$\triangle$FUV,
$\triangle$(FUV $-$ NUV)] = [0.01, 0.01] and reaching up to a maximum of [0.5, 0.5]. The number of sources present in the 
inner region and field region are 3880 and 2674 respectively. Following the above mentioned technique,
we removed 2507 number of
sources from the galaxy region as background contamination and are left with 1373 sources.
We have shown the CMD for 1373 sources in Figure \ref{cmd_clean}a. Since the fraction of background contaminants increases with decreasing brightness value, we have further excluded sources fainter than 23 mag in FUV band assuming them to be background galaxies. The foreground
stars are expected to
have a larger (FUV$-$NUV) color. In the study of extended disks of several nearby galaxies using GALEX data, Goddard et al. (2010) considered only sources with (FUV$-$NUV) $<$ 1.5 to exclude foreground and background sources. We treated sources with (FUV$-$NUV) $>$ 1.0 as not part of the galaxy and excluded them. The remaining sources are corrected for reddening and extinction (discussed in Section \ref{s_reddening}). The sources with corrected (FUV$-$NUV) $<$ $-$ 0.43 ($\sim$ 7\% of the remaining sources) are also excluded since their color value is outside the model range.
Finally, we are left with 742 sources in the galaxy region and considered them as candidate UV sources of the galaxy. The reddening and extinction corrected CMD for these 742 selected UV sources are shown in Figure \ref{cmd_clean}b. Since our main interest is to study the outer disk of the galaxy, out of these 742 UV sources we selected 261 sources, which are present between the radii 5.3 kpc (optical radius of NGC~300 (Faesi et al. 2014)) to 10 kpc, for our analysis. These 261 sources are shown in red in Figure \ref{cmd_clean}b.\\

We have also estimated the number of background galaxies in both GALEX FUV and NUV filters following the study of Xu et al. (2005). 
For the magnitude range 14.2 to 23.7, the estimated number of background galaxies are 0.7/arcmin$^2$ and 1.4/arcmin$^2$ 
respectively in FUV and NUV bands. Through our statistical method, we have excluded 2507 sources (
$\sim$ 93\% of the total field sources) from the 
galaxy region, which corresponds to a density of $\sim$ 1.2/arcmin$^2$, similar to the numbers provided above. 
We therefore conclude that our sample has negligible contamination due to background as well as foreground sources.\\

\section{Reddening and Metallicity}
\label{s_reddening}

We corrected reddening and extinction to estimate the intrinsic (FUV$-$NUV) color and both FUV and NUV magnitude for all selected sources. To study the outer 
stellar disk of NGC~300, Vlaji{\'c} et al. (2009) considered a foreground reddening E(B$-$V) = 0.011-0.014 mag.
The foreground reddening given in the GALEX catalog for the field of NGC~300 is also very low (E(B$-$V) $\approx$ 0.01). Apart from this, interstellar extinction also contributes to reddening inside the disk of a galaxy.
Using HST observation, Rodr{\'i}guez et al. (2016) studied different regions (from inner to outer disk) of NGC~300 by considering
a constant reddening of E(B$-$V) = 0.075 mag. In our study, we also adopted this reddening value throughout the galaxy.
Following 
conversion rules are used to calculate E(FUV$-$NUV), A(FUV) and A(NUV) from E(B$-$V).
\begin{equation}
E(FUV-NUV) = R(FUV-NUV)E(B-V)
 \end{equation}
 \begin{equation}
A(FUV) = R(FUV)E(B-V)  
 \end{equation}
 \begin{equation}
  A(NUV) = R(NUV)E(B-V)
 \end{equation}
We adopted the value of reddening coefficients from Yuan et al. (2013).
Using standard pair technique over a large sample of galactic stars, they estimated the values of 
R(FUV$-$NUV), R(FUV) and R(NUV) to be $-$2.35, 4.89 and 7.24 respectively for GALEX filters. 
The estimated values of E(FUV$-$NUV), A(FUV) and A(NUV) are $-$0.18 mag, 0.37 mag and 0.54 mag respectively.
These are used to correct color and magnitude of 
all the sources.
We used the reddening corrected color (FUV$-$NUV) and extinction corrected magnitude (FUV, NUV) throughout our analysis. In order to estimate different parameters of the detected sources, we adopted Z=0.02 as the present day metallicity of star forming regions from the study of supergiant stars in NGC~300 done by Gazak et al. (2015) and Kudritzki
et al. (2008).\\

The parameters of the galaxy are presented in Table 1. 
We calculated the inclination-corrected galactocentric distance (in kpc) for each source by considering the galaxy center as
$\alpha_{0}$ = 13.722833, $\delta_{0}$ = $-$37.684389, distance d = 1.9 Mpc, inclination i = $42.3^\circ$ and position angle (PA) of major
axis $\theta$ = $109^\circ$.\\

\section{SSP model}
\label{s_ssp}

We used starburst99 SSP model (Leitherer et al. 1999) to generate diagnostic diagrams
for studying the young star forming regions of the galaxy NGC~300. It is a spectrophotometric SSP model primarily used to understand the active star forming regions of a galaxy. Mondal et al. (2018) employed this model to estimate the masses of several compact star forming regions present in a dwarf irregular galaxy WLM. Faesi et al. (2014) 
used this model to estimate the mass and age of $H_{\alpha}$ regions present in the inner disk of NGC~300. 
This was also used by Goddard et al. (2010) to generate synthetic model grids. They compared these model grids 
with the GALEX FUV and NUV 
data of galaxy M31 and NGC 3621 to estimate mass and age of clusters. 
Dong et al. (2008) used starburst99 model to generate UV-MIR spectral energy distributions (SEDs) 
and obtained a constraint on the mass and age of clusters in the extended outer disk of M83.\\

\begin{table}
\centering
 
\caption{Starburst99 model parameters}
\label{starburst999}
\resizebox{90mm}{!}{
\begin{tabular}{cc}
\hline

 Parameter & Value\\\hline
 Star formation & Instantaneous\\
 Stellar IMF & Kroupa (1.3, 2.3)\\
 Stellar mass limit & 0.1, 0.5, 120 $M_{\odot}$\\
 Cluster mass range & $10^3 M_{\odot}$-$10^7 M_{\odot}$\\
 Stellar evolution track & Geneva (high mass loss)\\
 Metallicity & Z=0.04, 0.02, 0.008, 0.004, 0.001\\
 Age range & 1-900 Myr\\ \hline

\end{tabular}
}

\end{table}

We acquired starburst99 model data for the chosen set of parameters given in Table \ref{starburst999}.
Assuming the star formation to be instantaneous, we considered the IMF value as 1.3 (for 0.1 $M_{\odot}$ to 0.5 $M_{\odot}$), 2.3 (for 0.5 $M_{\odot}$ to 120 $M_{\odot}$)
(Kroupa IMF) (Kroupa 2001) with stellar mass limit as $M_{low}=0.1 M_{\odot}$ and $M_{up}=120 M_{\odot}$. We considered the total mass of the cluster in the range $10^3 M_{\odot}$-$10^7 M_{\odot}$. Nebular radiation for the total integrated light are also taken into account.\\

\subsection{Age}
\begin{figure}
\centering
\includegraphics[width=3.7in]{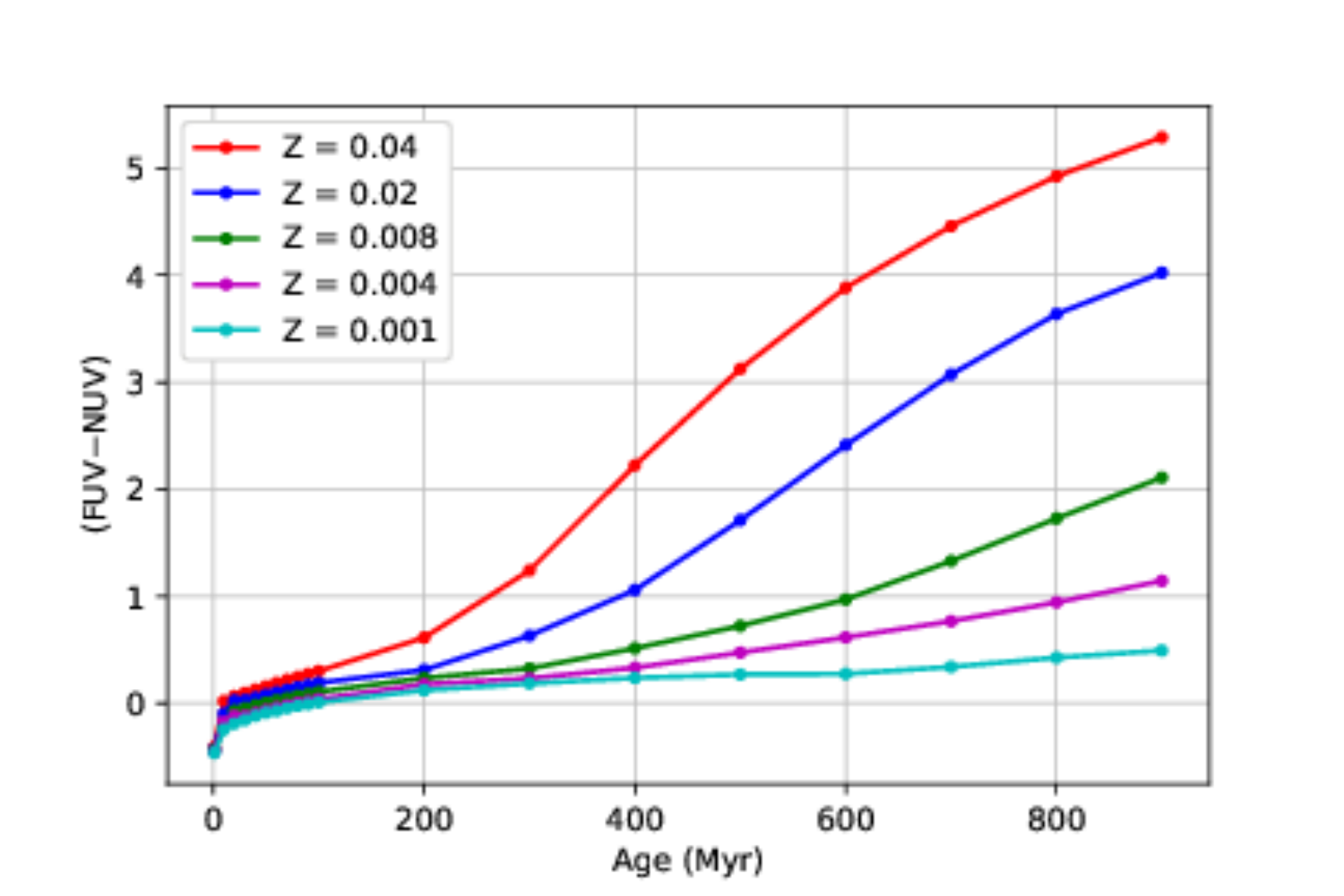} 
 \caption{Starburst99 model generated (FUV$-$NUV) color as a function of age (Myr). Different curves are for five 
 different metallicities 
(Z). For 1-100 Myr range, age interval is 10 Myr and after that interval is 100 Myr for age up to 900 Myr. The cluster mass considered is $10^6 M_{\odot}$.}
\label{GALEXmodel}
\end{figure}

We calculated (FUV$-$NUV) color for ages varying from 1 to 900 Myr with five different
metallicities (Z=0.04, 0.02, 0.008, 0.004, 0.001).
The diagnostic curves are shown in Figure \ref{GALEXmodel}.

Figure \ref{GALEXmodel} shows the following :\\
1. For a fixed metallicity, (FUV$-$NUV) color increases with age.\\
2. Change in color for the entire age range increases with metallicity.\\
For a given metallicity, we will be able to calculate the age of any source from its (FUV$-$NUV) color with the help of this 
model grid. We have considered the age range as 1-900 Myr. It should be noted that (FUV$-$NUV) color is not 
sensitive to estimate age beyond $\sim$ 500 Myr because of the substantial drop in the far-UV flux. \\

\subsection{Mass}
\begin{figure}
\centering
\includegraphics[width = 3.7in]{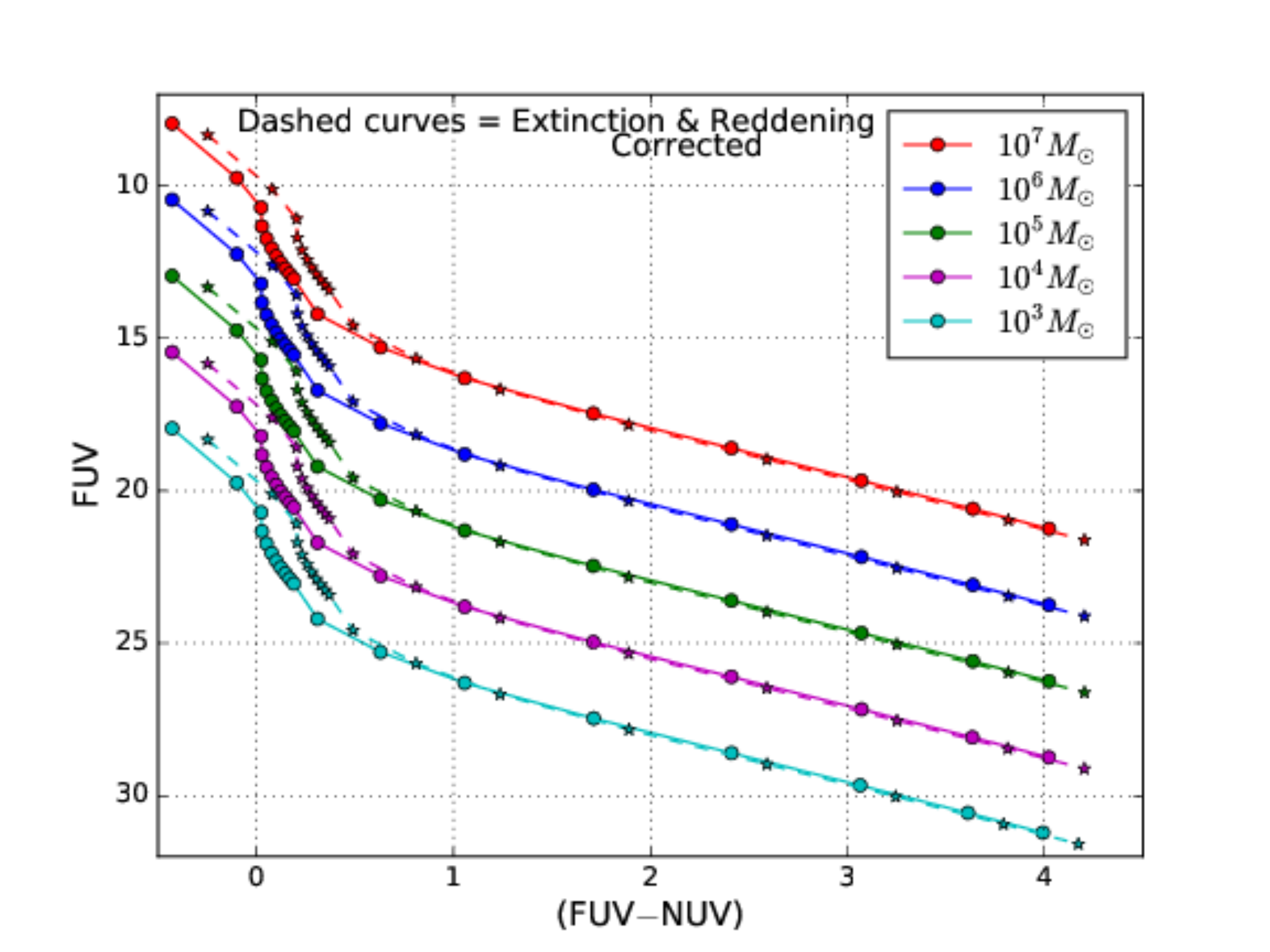}
 \caption{Starburst99 model generated FUV vs (FUV$-$NUV) CMD for simple stellar population. Different curves signify five different
total cluster mass ($10^7 M_\odot$, $10^6 M_\odot$, $10^5 M_\odot$, 
$10^4 M_\odot$, $10^3 M_\odot$). The points shown in each curve are for different ages starting from 1 Myr to 900 Myr 
(increasing along the color axis) with the same age interval chosen in Figure \ref{GALEXmodel}.}
    \label{cmd_model}
\end{figure}

We also produced model grids for FUV magnitude as a function of (FUV$-$NUV) color. 
The model fluxes and the corresponding magnitudes are estimated by adopting a distance of 1.9 Mpc for NGC~300 (Rizzi et al. 2006).
Keeping all model parameters 
unchanged, we considered five different values for the total cluster mass ($10^7 M_\odot$, $10^6 M_\odot$, $10^5 M_\odot$, 
$10^4 M_\odot$, $10^3 M_\odot$) for a fixed metallicity of Z=0.02, which is adopted for NGC~300,
to generate Figure \ref{cmd_model}.

Figure \ref{cmd_model} shows the following :\\
1. Older clusters have redder (FUV$-$NUV) color irrespective of mass.\\
2. For a given age (or equivalently, color), clusters become more brighter with increasing mass.\\
These model grids can be used to calculate both age and mass of sources from their color and magnitude for Z=0.02.\\

\section{Analysis}
\label{s_analysis}
\subsection{FUV Disk of NGC~300}

\begin{table}
\centering
\footnotesize
 
\caption{Details of flux and magnitudes for contours in the FUV map shown in Figure \ref{fuv_mass}.}
\label{FUV}
\begin{tabular}{p{3cm}p{3cm}p{1cm}}
\hline
FUV magnitude range (extinction corrected) & $Flux_{FUV}$ ($erg/sec/cm^2/\AA$) (extinction uncorrected) &  Contour color\\\hline
$<$ 20 & $> 33.6\times10^{-15}$ & Blue\\
$>$ 20 \& $<$ 21 & $13.3-33.6 \times10^{-15}$ & Green\\
$>$ 21 \& $<$ 23 & $0.21-13.3 \times10^{-15}$ & Red\\\hline
\end{tabular}

\end{table}

\normalfont

\begin{figure}
\begin{center}
\includegraphics[width=3.3in]{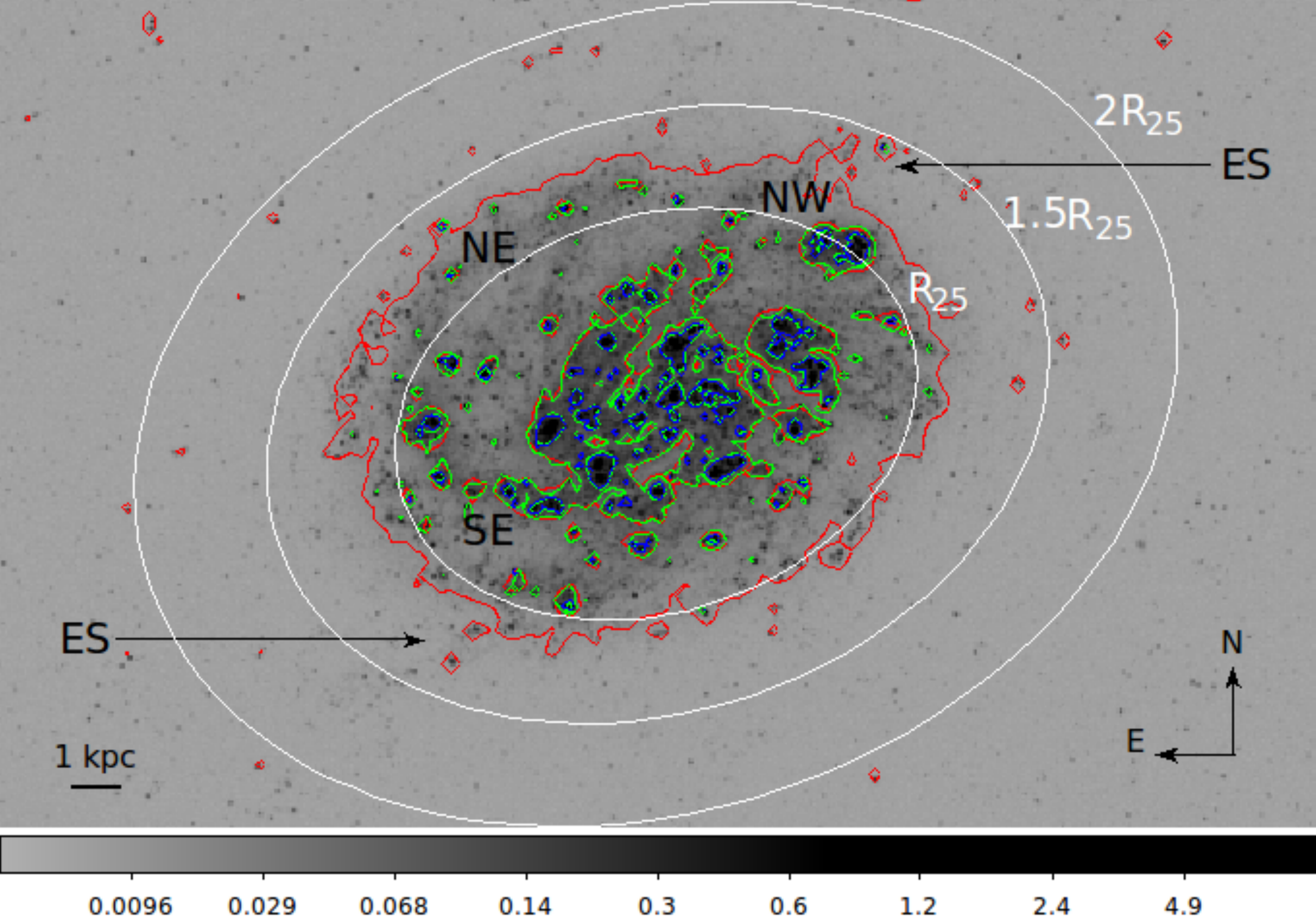} 
 \caption{The background figure is the binned $960\times960$ pixel FUV image of NGC~300 with different contours plotted for different limits of FUV flux value as mentioned in Table \ref{FUV}. The blue contours represent the brightest (hence massive) regions of the galaxy. The gray scale of the image denotes counts per second per pixel. The ellipses shown in the figure signify $R_{25}$, 1.5$R_{25}$ and 2$R_{25}$ galactocentric distance respectively. Two extended structures (ES) are also shown by arrow.}
 \label{fuv_mass}
 \end{center}
 \end{figure}

In order to identify large massive star forming complex of NGC~300 we produced a 960$\times$960 pixel image from the 3840$\times$3840 pixel GALEX FUV image by binning 4$\times$4 pixels of original image as one single pixel. Each pixel of the binned image has an area coverage of $\sim$ 6$\times$6 arcsec$^2$ ($\sim$ 55$\times$55 pc$^2$). The resolution of the image is degraded such that the pixel size becomes comparable to the average size of young stellar groups in NGC~300 identified by Rodr{\'i}guez et al. (2016). Since far-UV radiation is mainly contributed by massive young stars, we considered the image in FUV filter for this purpose. To understand the distribution of star forming regions, we created different contours by fixing the lower and upper limit of fluxes which are given in Table \ref{FUV}. The FUV image with different contours is shown in Figure \ref{fuv_mass}. The massive star forming complexes of NGC~300 are picked up by the blue contours created for pixel brighter than 20 magnitude (i.e. $Flux_{FUV}>33.6\times10^{-15}$ $erg/sec/cm^2/\AA$).
The two main spiral arms of the galaxy, one along north-west (NW) and another along east to north-east (NE), are found to have massive star forming complexes which all together follow the spiral structure. Almost all of these complexes are found to be present within the optical radius ($R_{25}$) of the galaxy. The green contours, signifying relatively less bright regions, are found around each of the blue contoured regions. The red contours, generated for pixels fainter than 21 magnitude and brighter than 23 magnitude, trace the extended structure of the FUV disk nearly up to a radius 7 kpc. The north-eastern spiral arm, which do not have much massive complexes, is mainly covered by this red contour. We do see two extended structures along the major axis (south-east (SE) to NW) of the galaxy which are mainly picked up by the red contours. The regions covered by red contour in the east and northern part, extend beyond the optical radius ($R_{25}$) of the galaxy. Therefore, NGC~300 has an extended UV disk with relatively less massive star forming complexes when compared to the inner disk.\\

\subsection{Correlation with optical image}
\begin{figure}
\begin{center}
\includegraphics[width=3.3in]{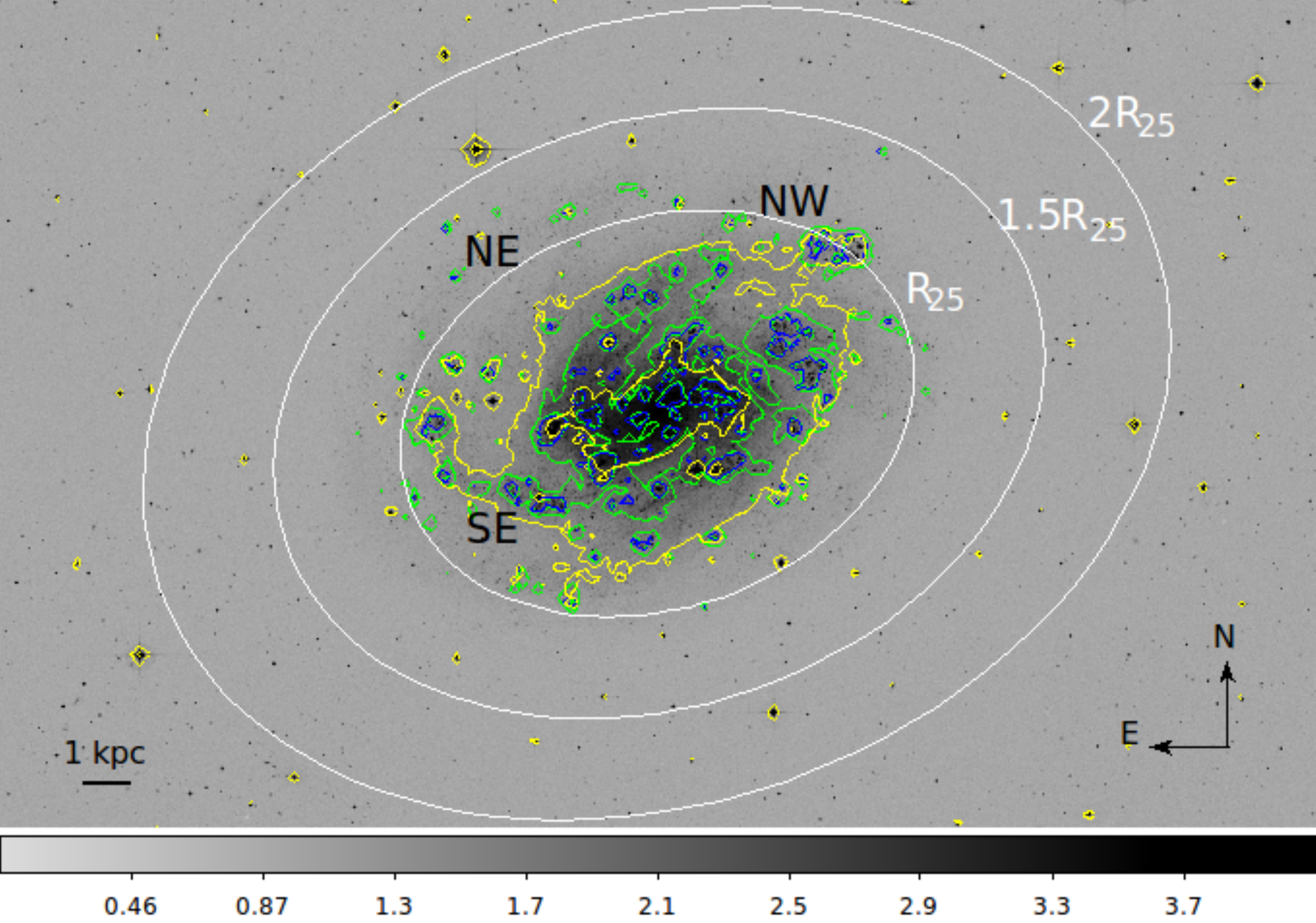} 
 \caption{The background figure shows the DSS optical image of NGC~300. The yellow contoured region represents the main optical disk of the galaxy. Blue and green contours are same as shown in Figure \ref{fuv_mass}. }
 \label{dss}
 \end{center}
 \end{figure}

As the young massive stars have a large continuum flux in UV, when compared to optical, a correlation between UV and optical emission will eventually help to trace the young star forming regions.
In order to correlate the FUV disk of NGC~300 with the optical disk, we have over plotted the blue and green contours of Figure \ref{fuv_mass} on the DSS optical image of the galaxy in Figure \ref{dss}. The yellow contour in Figure \ref{dss} displays the main part of the optical disk of the galaxy. It is noticed that the massive complexes identified in the GALEX FUV image of NGC~300 (blue and green contour) correlate well with the optical disk structure (yellow contour) of the galaxy. It is to be noted that, the optical emission arising from the north-eastern spiral arm of the galaxy is very low compared
to other regions of the galaxy, whereas in the FUV image (Figure \ref{fuv_mass}), we can clearly notice the spiral arms and the extended structures present in the outer disk of the galaxy. Therefore, the extended disk, identified in UV, is not easily noticeable in the optical image of the galaxy.\\  

\subsection{Correlation with H~I}

\begin{figure}
\centering
\subfigure[]{\includegraphics[width = 3.5in]{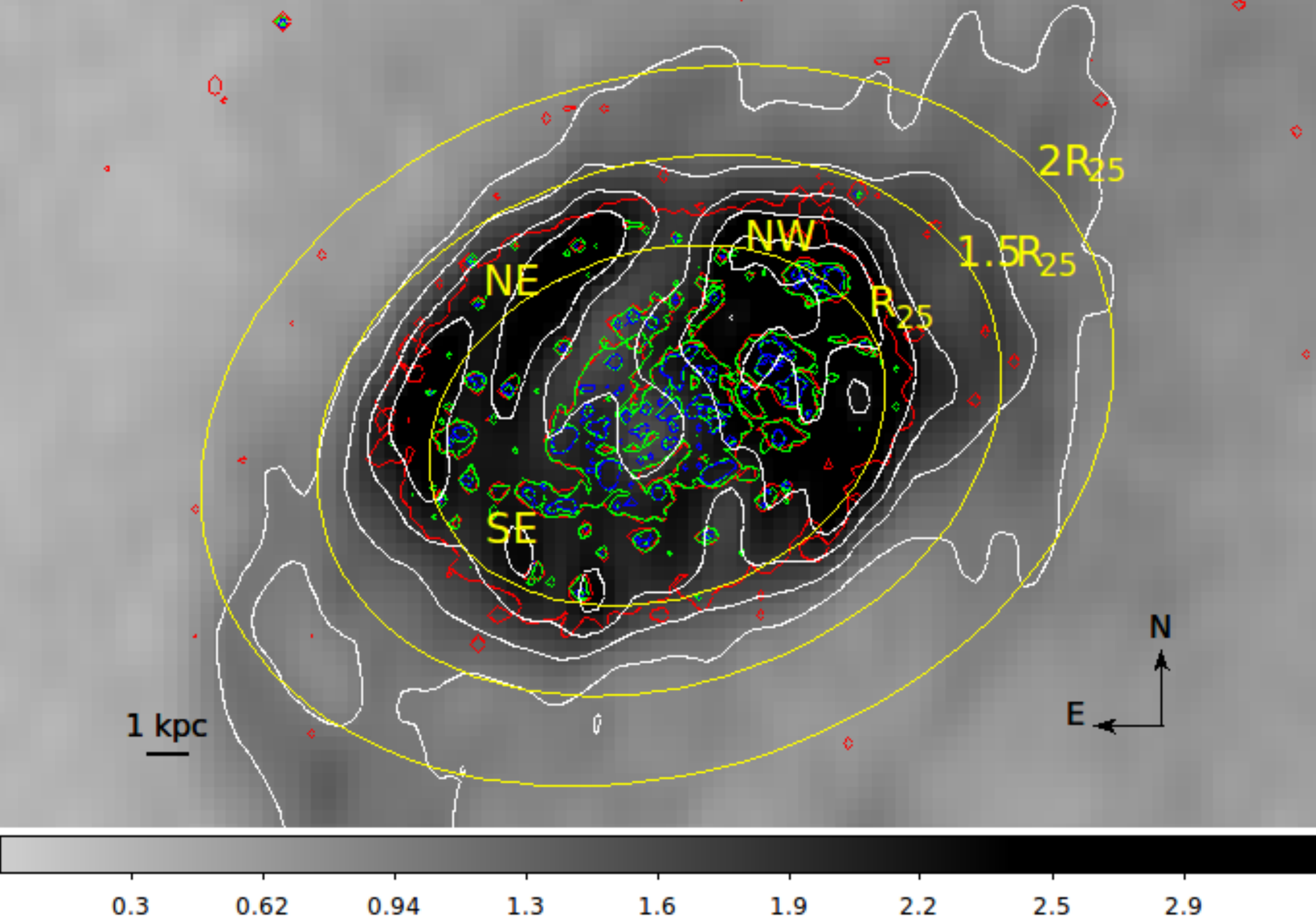}}
\subfigure[]{\includegraphics[width = 3.5in]{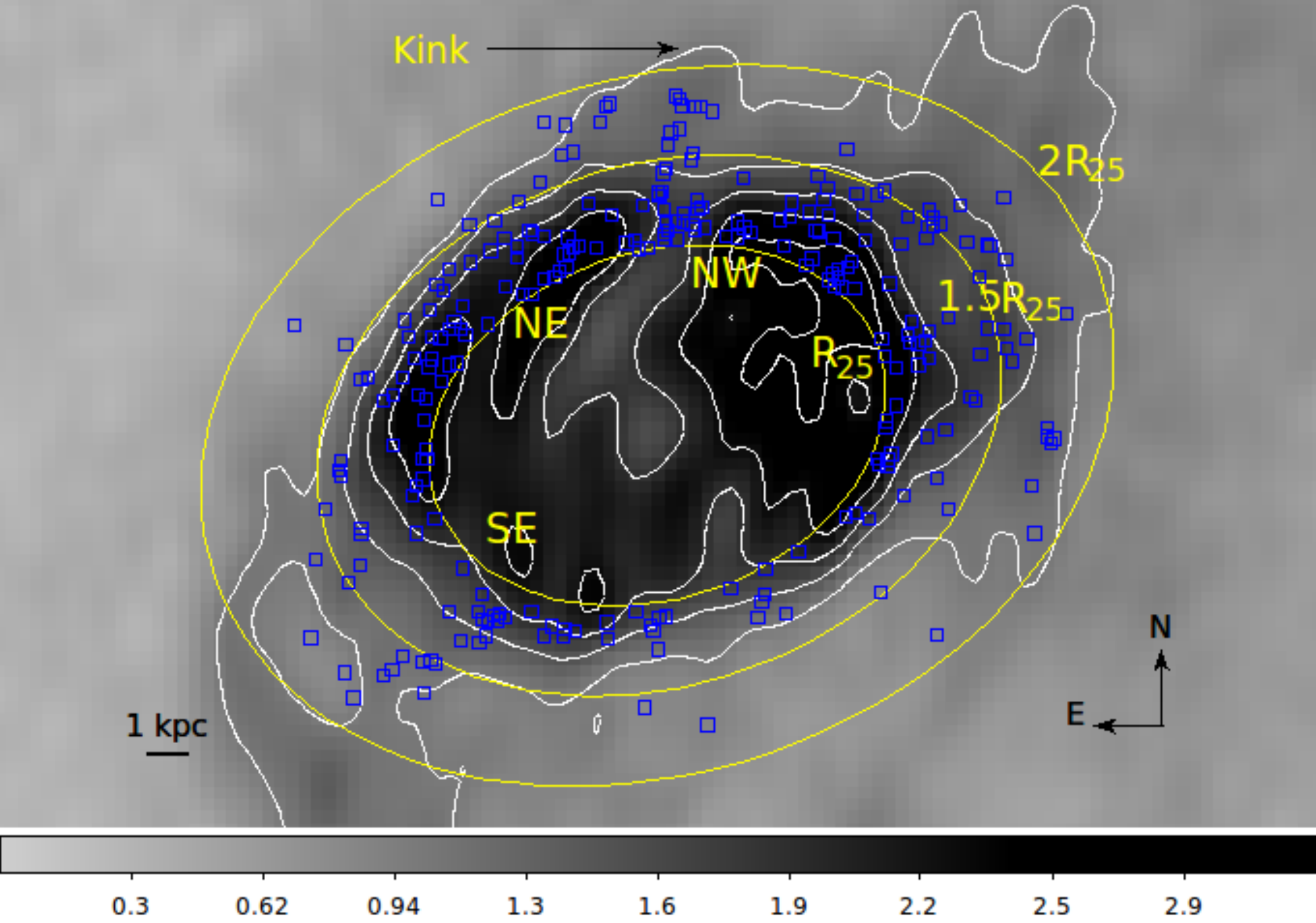}}
 \caption{The H~I density contours are shown in white on the background H~I density map of NGC~300 in both figures. The unit of the gray scale is Jy/beam. The blue, green and red contours shown in Figure (a) are the same as Figure \ref{fuv_mass}. In Figure (b) we have shown the distribution of UV sources (blue points) present in the outer disk (between 5.3 kpc and 10 kpc) of NGC~300.}
 \label{HI_distribution}
 \end{figure}

In order to correlate the distribution of young star forming regions and H~I column density of the galaxy, we over plotted the FUV contours of Figure \ref{fuv_mass} on the H~I map in Figure \ref{HI_distribution}a. The H~I density contours, generated for a higher threshold value, are shown in white in the same figure. It is noticed that the extent of the detected FUV disk of the galaxy is well within the dense H~I disk. 
 H~I contour map shows a large compression of gas in the south-eastern part of the galaxy. Westmeier
et al. (2011) found that the south-eastern side showed a distinctive plateau, followed by a sudden and steep drop in the column density of the H~I disk, which is not found in the north-western side. The UV sources which are present along the north-eastern spiral arm of the galaxy are also found to coincide with the dense H~I contours
(Figure \ref{HI_distribution}b). A kink in the density contour is also seen in the northern direction and 
it coincides with the extended structure identified in the source distribution. The extended features noticed along south-east and north-west direction also follow the extended density contour of H~I.\\

\subsection{Correlation with 24 $\mu$m infrared image}
\begin{figure}
\begin{center}
\includegraphics[width=3.3in]{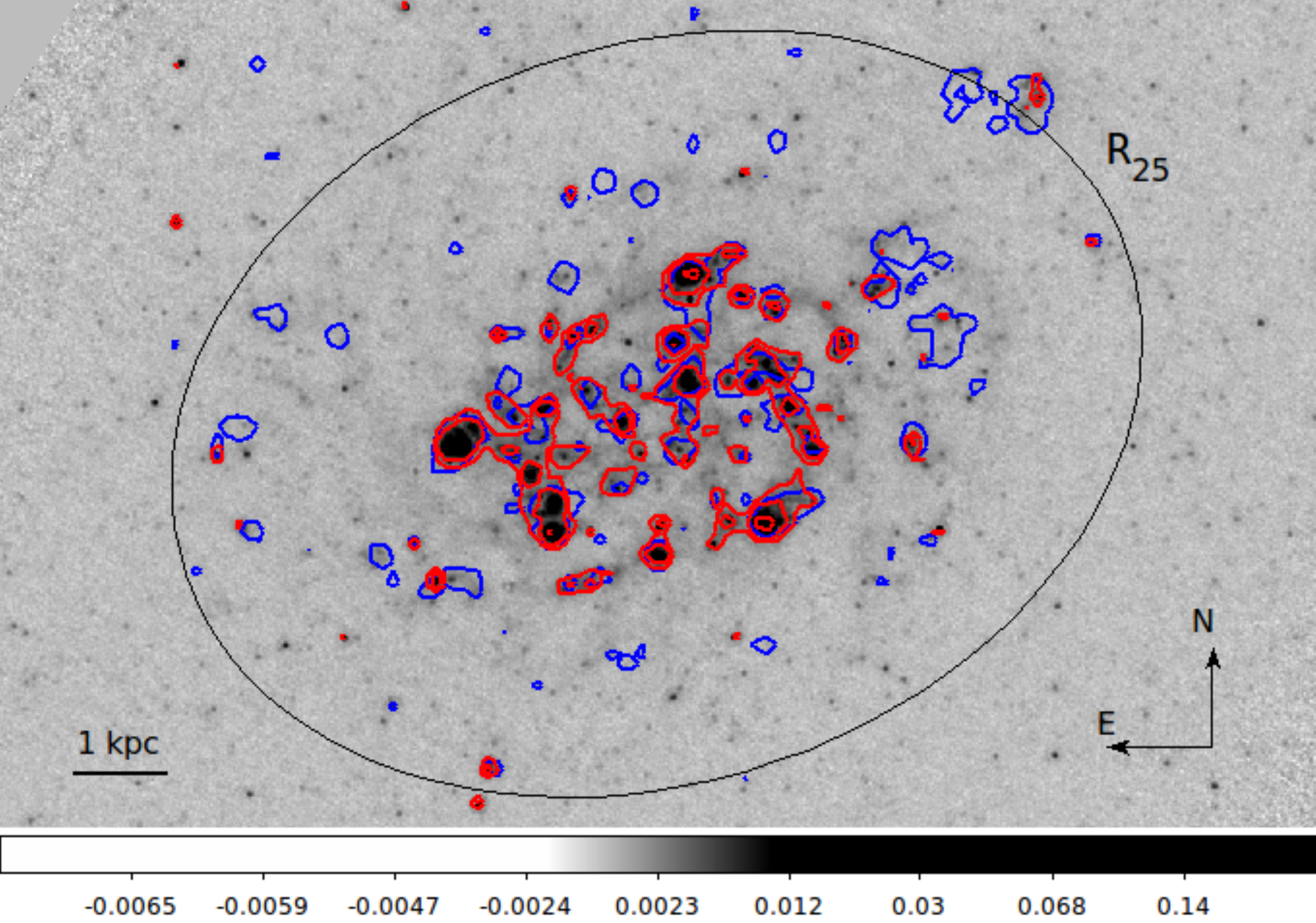} 
 \caption{The MIPS 24 $\mu$m image of NGC~300 is shown in the background with red contours indicating regions with intense infrared emission. The same blue contours of Figure \ref{fuv_mass} are also shown. The unit of the gray scale is MJy/sr.}
 \label{ir}
 \end{center}
 \end{figure}

The star forming regions of the galaxy NGC~300 are found to have strong 24 $\mu$m infrared emission (Helou et al. 2004). Since the dust present in the star forming regions gets heated by the FUV photons originating from the massive stars which in turn produces radiation in 24 $\mu$m, it is expected that the massive star forming complexes detected in FUV image of NGC~300 should also show emission in 24 $\mu$m. In order to verify this, we over plotted the blue contour (of Figure \ref{fuv_mass}), signifying massive star forming complex, on the infrared 24 $\mu$m image of the galaxy in Figure \ref{ir}. The red contours shown in the figure indicate the regions with intense 24 $\mu$m emission. It is noticed that for the inner disk both the blue and red contours show a good spatial correlation which signifies the location of active star forming regions in NGC~300. We also note that a few FUV emission regions located within $R_{25}$, do not have corresponding 24 $\mu$m emission. This may be due to a variety of reasons such as, difference in resolution and sensitivity of the detections in these wavelengths, reduced dust in the outer regions, presence of a few possible not-so-young star forming regions without dust etc. \\

\subsection{Luminosity density profile}

\begin{figure}
\begin{center}
\includegraphics[width=3.5in]{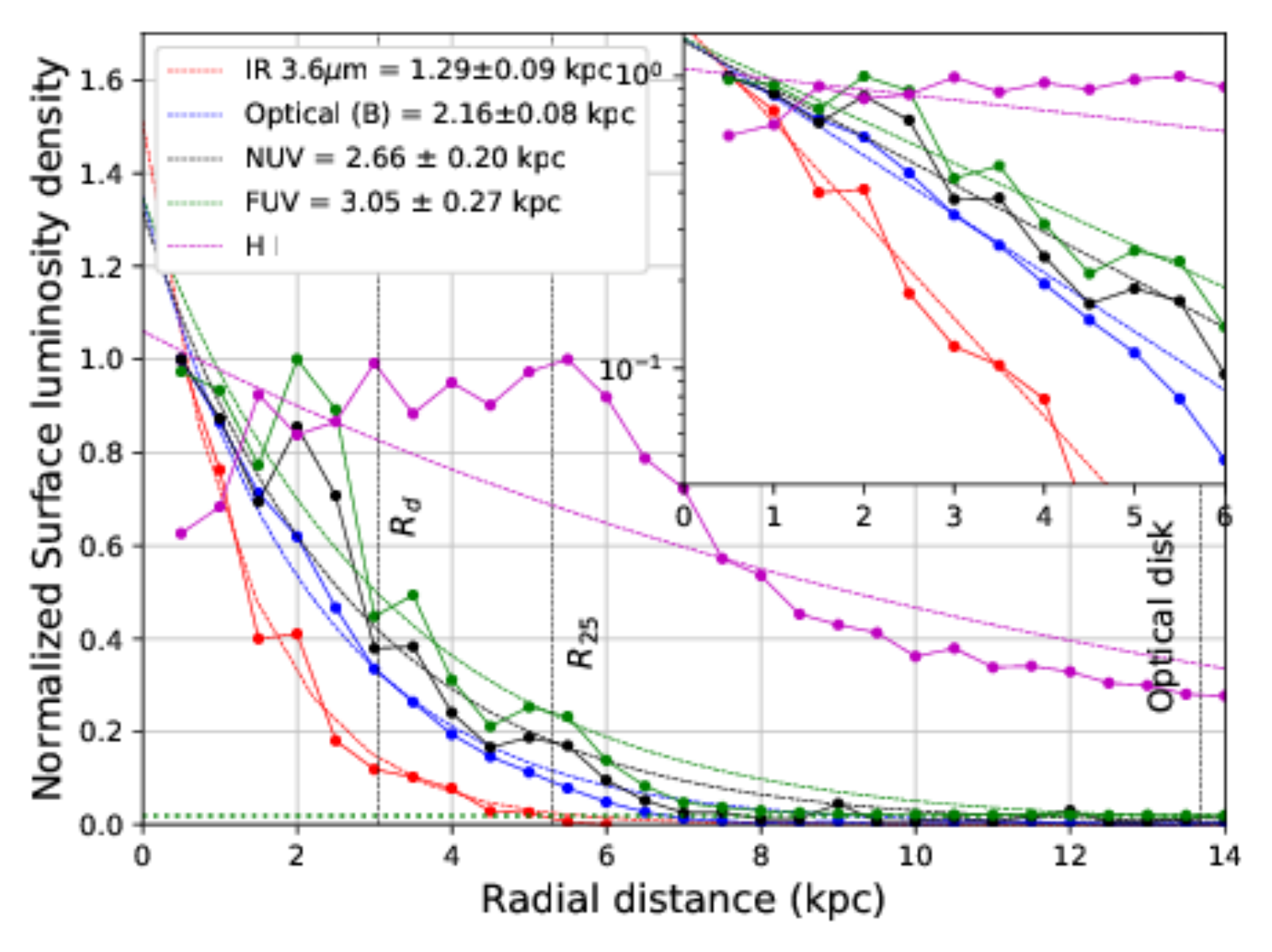} 
 \caption{The normalized surface luminosity density profiles of NGC~300 in different wavebands are shown in different colors. The solid lines denote the observed profiles whereas dashed lines of the same color show the fitted exponential profiles. Three vertical black dashed lines are plotted to show the FUV scale-length ($R_d$) (our study), optical radius ($R_{25}$) and the extend of optical disk of the galaxy from Bland-Hawthorn et al. 2005. The horizontal dotted green line represents the FUV background. In inset, we have shown all the observed and fitted profiles in logarithmic scale up to radial distance of 6 kpc.}
 \label{luminosity}
 \end{center}
 \end{figure}

The emission from stellar population of different ages peaks in different wavelengths. The older populations show significant emission in optical and infrared bands whereas emission from younger populations mainly peak in UV. In order to trace the disk of NGC~300 in different wavelengths, we produced normalized surface luminosity density profiles of the galaxy in different wavebands which are shown in Figure \ref{luminosity}. These profiles depict the nature of the galaxy disk in different wavelengths. The disk scale-length ($R_d$) of NGC~300 is reported to be 2.10 kpc and 1.47 kpc respectively in optical B band and infrared I band (Carignan et al. (1985), Kim et al. (2004)). These two independent measurements highlight that the disk is more extended in shorter wavelengths. We used GALEX FUV and NUV, DSS optical B band and infrared 3.6 $\mu$m IRAC images to estimate disk scale-length by fitting an exponential curve to each of the observed profiles, shown in Figure \ref{luminosity} in different colors. The measured values of scale-length are 1.29$\pm$0.09 kpc, 2.16$\pm$0.08 kpc, 2.66$\pm$0.20 kpc and 3.05$\pm$0.27 kpc respectively for infrared, optical B band, NUV and FUV respectively. This suggests that the disk of NGC~300 gradually extends to larger radii from longer to shorter wavelength. The distribution of younger populations, traced by the FUV disk, is found to be more extended than the rest. This thus confirms the presence of XUV disk in the galaxy NGC~300. The average FUV background, estimated from the flux measured between radii 14 and 15 kpc, is shown in green dotted line. The observed FUV luminosity density profile nearly converges to background level at radius $\sim$ 12 kpc. Therefore, we conclude that the disk of NGC~300 is extended at least up to $\sim$ 12 kpc. We have also shown the radial H~I column density profile of the galaxy in the same figure. The behaviour of this profile is found to be different than the rest. Instead of an exponential nature like other bands, it shows a dip in the central part with gradual increase up to a radius $\sim$ 5.5 kpc. Beyond this radius the profile drops down slowly in the outer disk. We did not show the H$\alpha$ profile because the image available for that in NED has a radial coverage of around 4 kpc, which is less than the optical radius of the galaxy.

 \begin{figure}
\begin{center}
\includegraphics[width=3.7in]{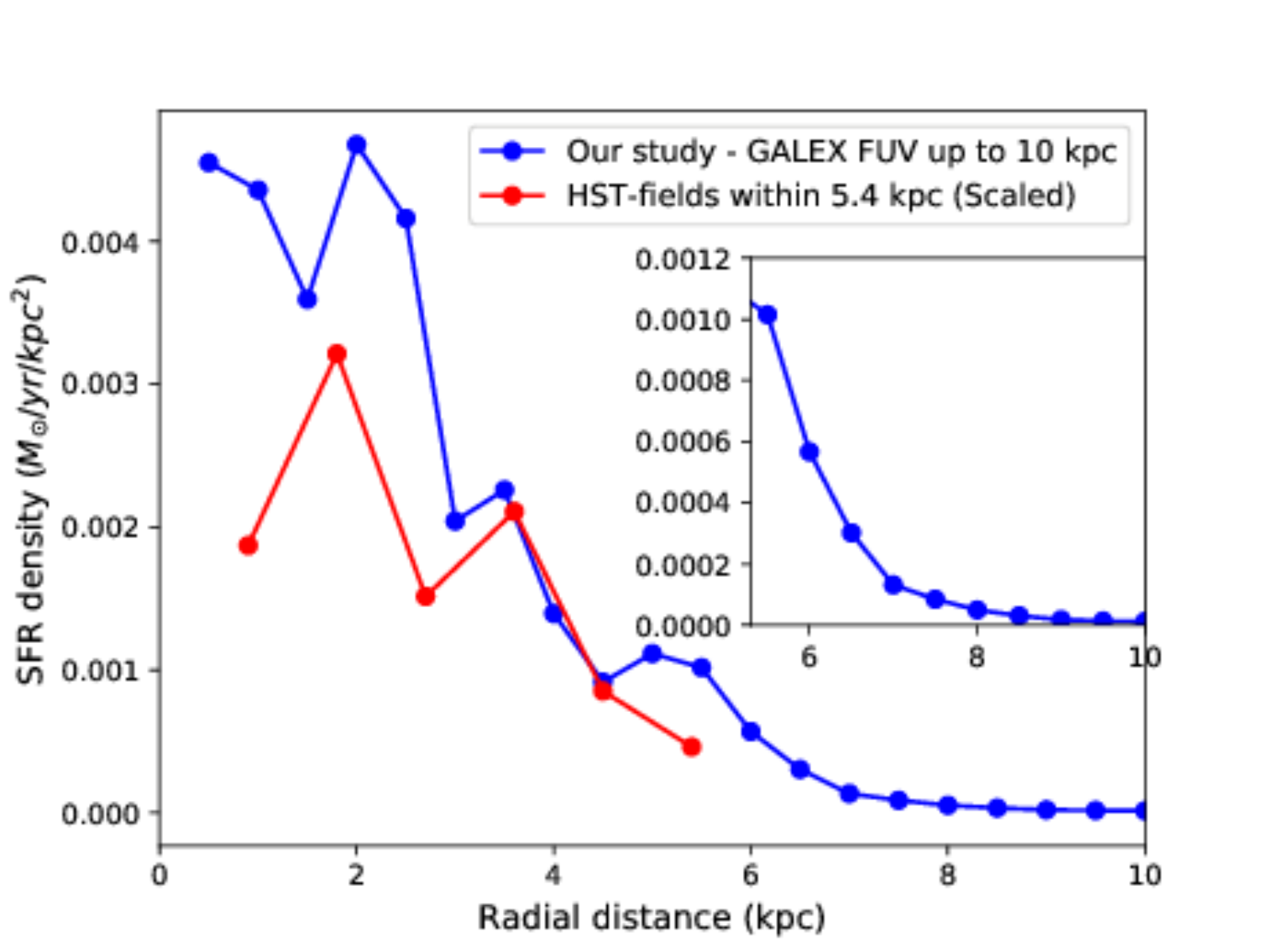}
 \caption{The radial profile of SFR density ($M_{\odot}/yr/kpc^2$) is shown up to a radius 10 kpc (blue curve). The red curve shows the scaled value of SFR density ($M_{\odot}/yr/kpc^2$) as calculated through an HST study by Gogarten et al. (2010). The radial profile of SFR density between radii 5.3 kpc and 10 kpc is shown in the inset.}
 \label{sfr_radial}
 \end{center}
 \end{figure}
 
 \subsection{Star formation rate}
To estimate the SFR of NGC~300, we considered a radius of 10 kpc (1 $\sim$ 2 R$_{25}$) and measured the total flux within that from the GALEX FUV image. This flux is corrected for extinction and background as discussed in previous section and further converted to magnitude. We used equation \ref{sfr_eq} from Karachentsev \& Kaisina (2013), where $mag_{FUV}$ denotes the background and extinction corrected magnitude and D is the distance to the galaxy in Mpc, to calculate the SFR ($M_{\odot}/yr$) of the galaxy. The total SFR integrated up to a radius 10 kpc for NGC~300 is found to be 0.46 $M_{\odot}/yr$, which nearly matches with the earlier estimate of 0.30 $M_{\odot}/yr$ by Karachentsev \& Kaisina (2013). To generate the radial profile of SFR density ($M_{\odot}/yr/kpc^2$), we similarly considered annuli of width 0.5 kpc from the center to a radius 10 kpc and measured the background and extinction corrected FUV magnitude per kpc$^2$ in each individual annuli. We used equation \ref{sfr_eq} to estimated the SFR for each annuli and plotted radially in Figure \ref{sfr_radial}. The profile shows that the central 2 kpc of the galaxy shows high SFR density. After that it starts decreasing and becomes little stable inside the radius 5.5 kpc. Beyond this radius the SFR density becomes very small compared to the inner disk. In a study of NGC~300 using HST data, Gogarten et al. (2010) calculated the radial SFR density ($M_{\odot}/yr/kpc^2$) for different age ranges. They used three HST ACS fields oriented as a radial strip from center to outside of the galaxy in the western part. These fields thus cover a limited part of the  galaxy's inner disk. Our study instead considered the whole galaxy disk observed  by GALEX to estimate the azimuthally averaged radial SFR density profile of NGC 300.  Both of these measurements have different area coverage and hence values are expected to be different. In order to check only the nature of radial profile from both the studies, we have plotted the values from Gogarten et al (2010) for the age range 4-80 Myr (Figure \ref{sfr_radial} (red curve)), after scaling it to our estimations. The plot highlights that though HST covers a specific part of the disk, the nature of radial profile closely follows that estimated for the whole disk with GALEX in this study.\\
 
 \begin{equation}
log(SFR_{FUV} (M_{\odot}/yr)) = 2.78 - 0.4*mag_{FUV} + 2log(D) 
\label{sfr_eq}
 \end{equation}

 \subsection{Age estimation of UV sources}
\begin{figure}
\centering
\includegraphics[width=3.5in]{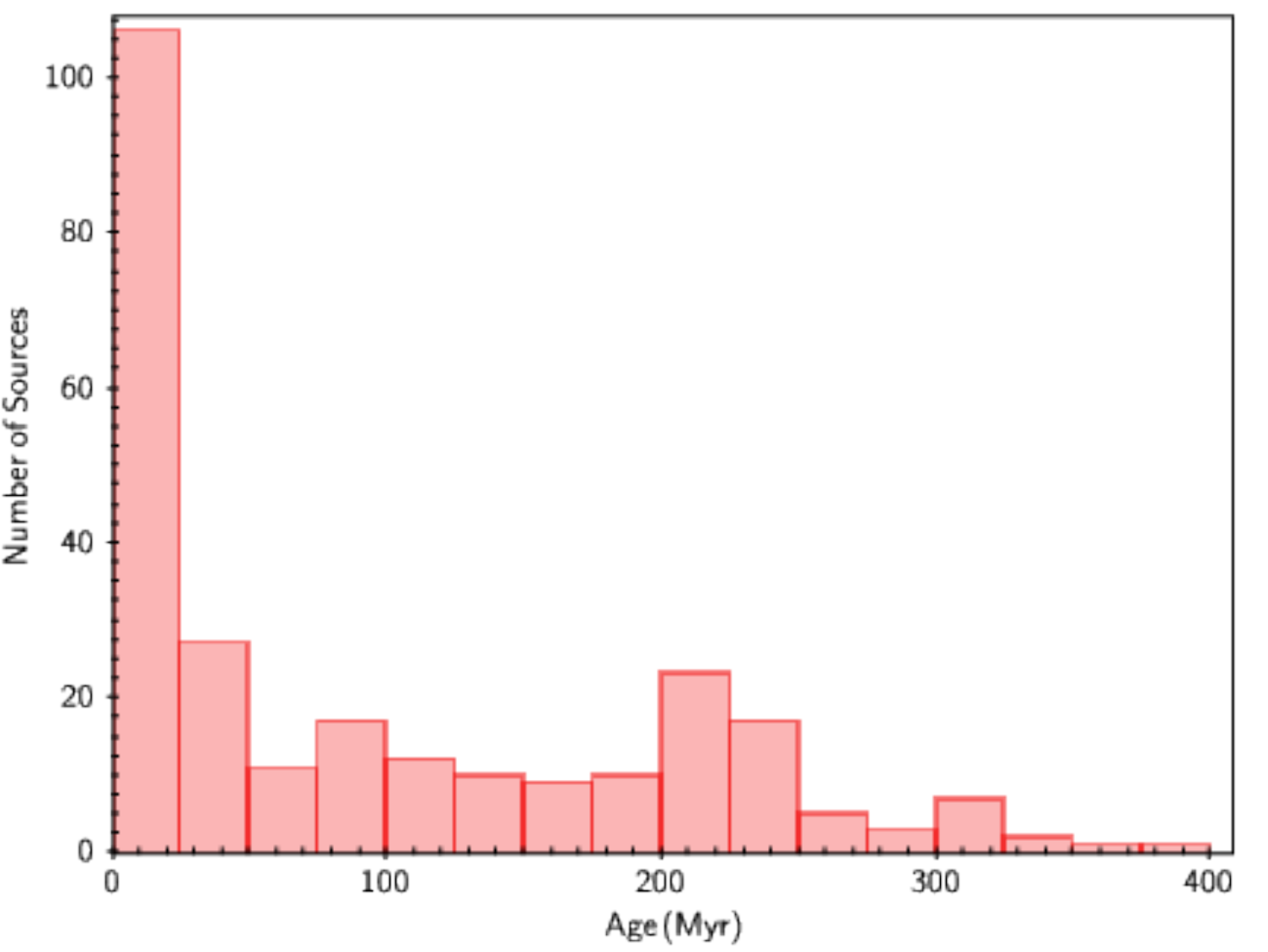} 
 \caption{Age histogram of all the selected UV sources between radius 5.3 kpc and 10 kpc.}
\label{age_hist}
\end{figure}

The diagnostic diagram shown in Figure \ref{GALEXmodel} can be used to estimate the age of a source from its (FUV$-$NUV)
color for a given metallicity. As NGC~300 is observed to have solar metallicity, we considered the
blue curve (Z=0.02) in Figure \ref{GALEXmodel} to estimate the age of identified UV sources. The sources detected in the inner disk have crowding issue. Because of the poor spatial resolution of GALEX, a combination of multiple sources in the inner disk can appear like a single source. Whereas in the outer disk of the galaxy, artefact due to crowding is much reduced. Also, out of 742 we selected 261 sources having galactocentric distance greater than 5.3 kpc and less than 10 kpc for our study. We do not consider sources beyond 10 kpc for contaminants might be more in the far outer disk. We used the reddening corrected observed color of these 261 UV sources and interpolated it with the model data (blue curve) to estimate their ages within the range 1 to 400 Myr. In Figure \ref{age_hist}, we have shown the age histogram for these selected UV sources. The figure signifies that the outer disk of NGC~300 has a large number of young sources with age less than 25 Myr.\\

\begin{figure}
\centering
\subfigure[]{\includegraphics[width = 3.2in]{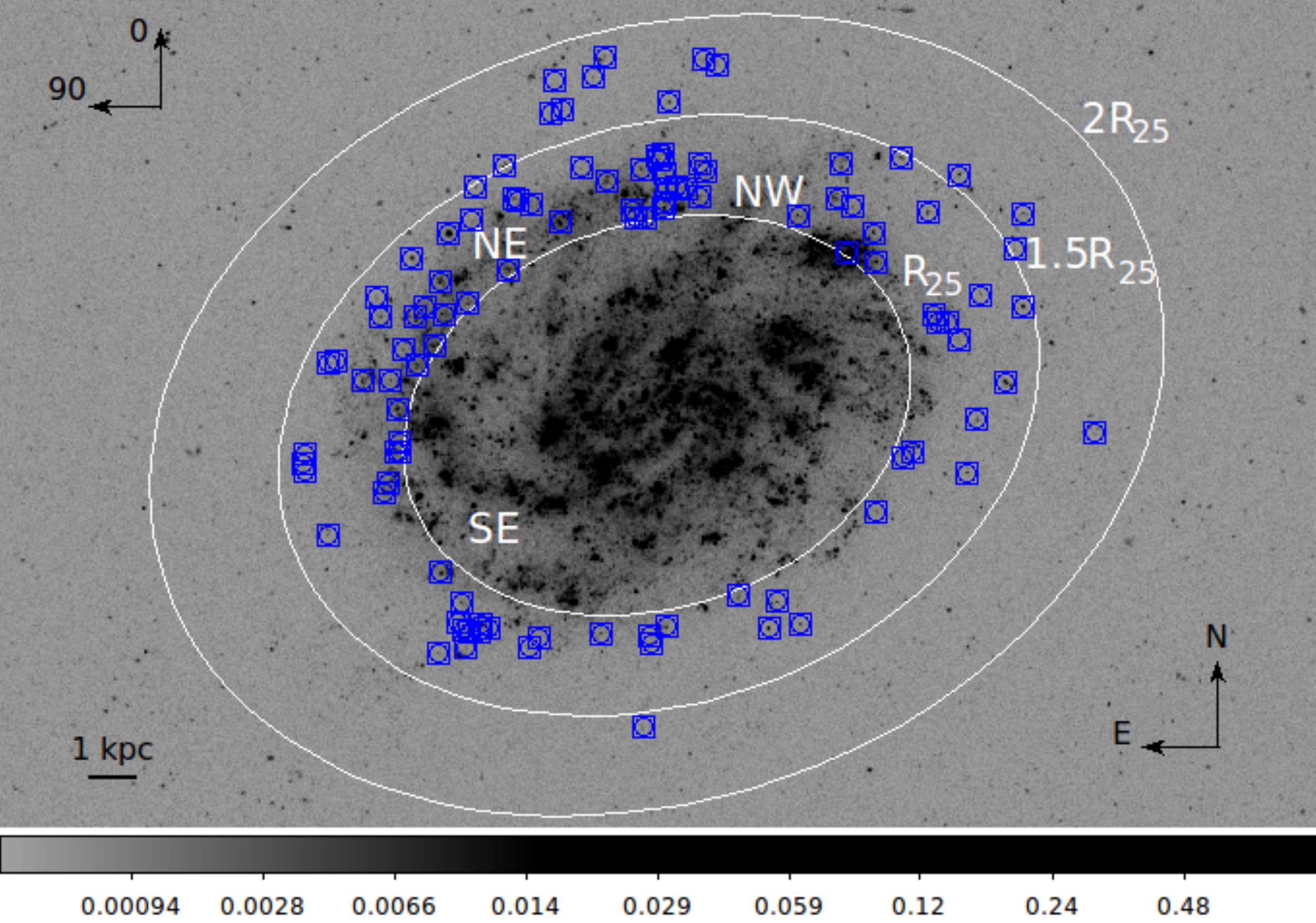}}
\subfigure[]{\includegraphics[width = 3.2in]{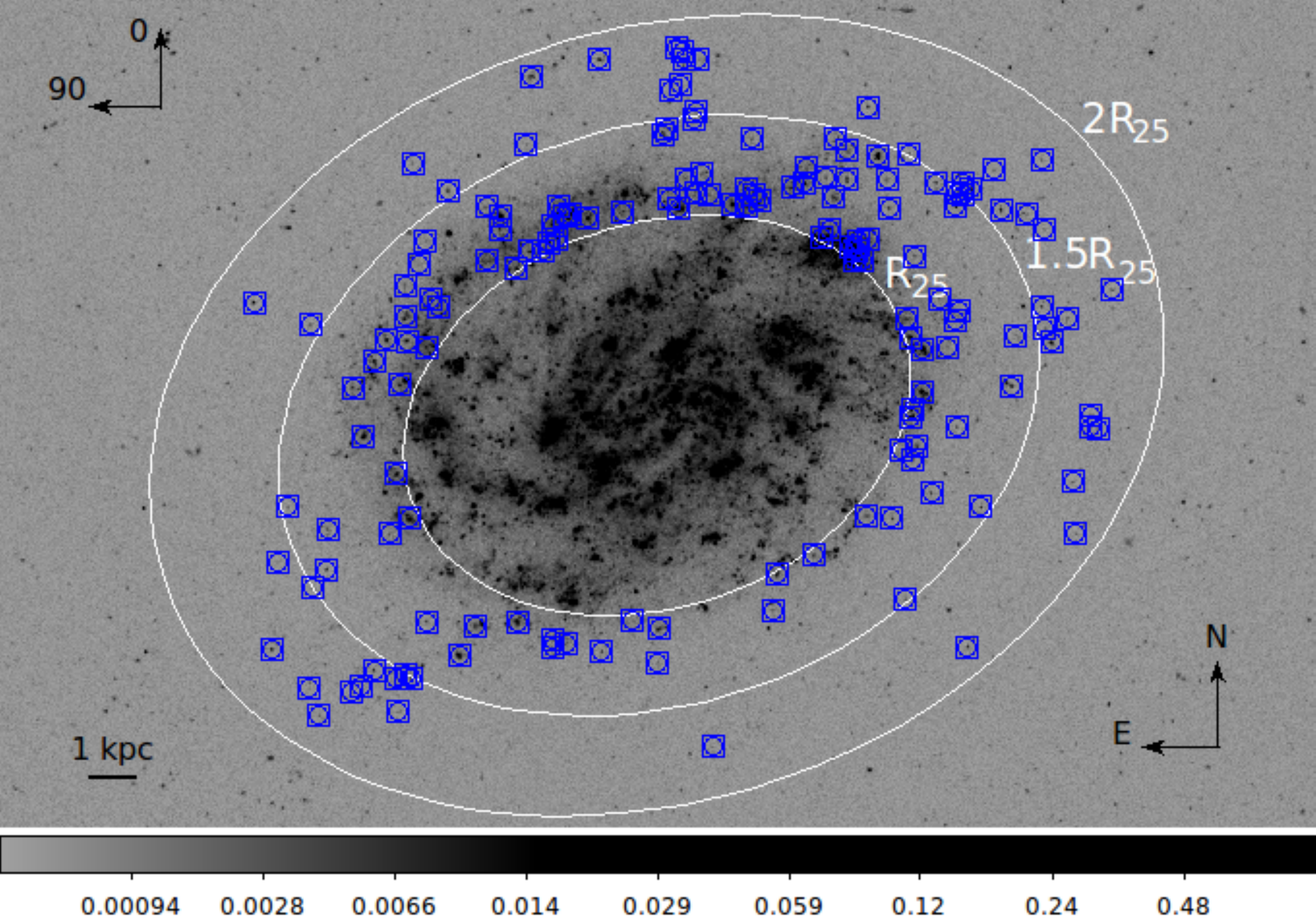}}
 \caption{The selected UV sources (blue square) in the outer disk are over-plotted on the GALEX FUV image of the galaxy. Figure (a) shows the spatial position of younger sources (Age $<$ 25 Myr) and the relatively older sources (Age $>$ 25 Myr) are shown in Figure (b). The reference PA is shown in the top left corner.}
 \label{age_dist}
 \end{figure}
 
\subsection{Spatial age distribution of UV sources}
In order to visualize the spatial age distribution of these 261 selected UV sources, we separated them into two groups, one as the young group (Age $<$ 25 Myr) and another one as relatively older group (Age $>$ 25 Myr). The sources in each groups are then over plotted on the GALEX FUV image in Figure \ref{age_dist}. 60\% of the sources with age $<$ 25 Myr are found to be present between PA 0 - 180$^\circ$ whereas 40\% are present between PA 181-360$^\circ$. In case of sources older than 25 Myr, the scenario is exactly opposite to this. Majority of the younger sources (Age $<$ 25 Myr) are found to be present along the north-eastern spiral arm of the galaxy (Figure \ref{age_dist}a). The extended part of a spiral arm seen in the western part of the galaxy are found to be populated by most of the older sources (Figure \ref{age_dist}b). The elongated distribution of sources along the major axis of the galaxy (south-east to north-west) is seen only for the older sources.\\

 \begin{figure}
\begin{center}
\includegraphics[width=3.7in]{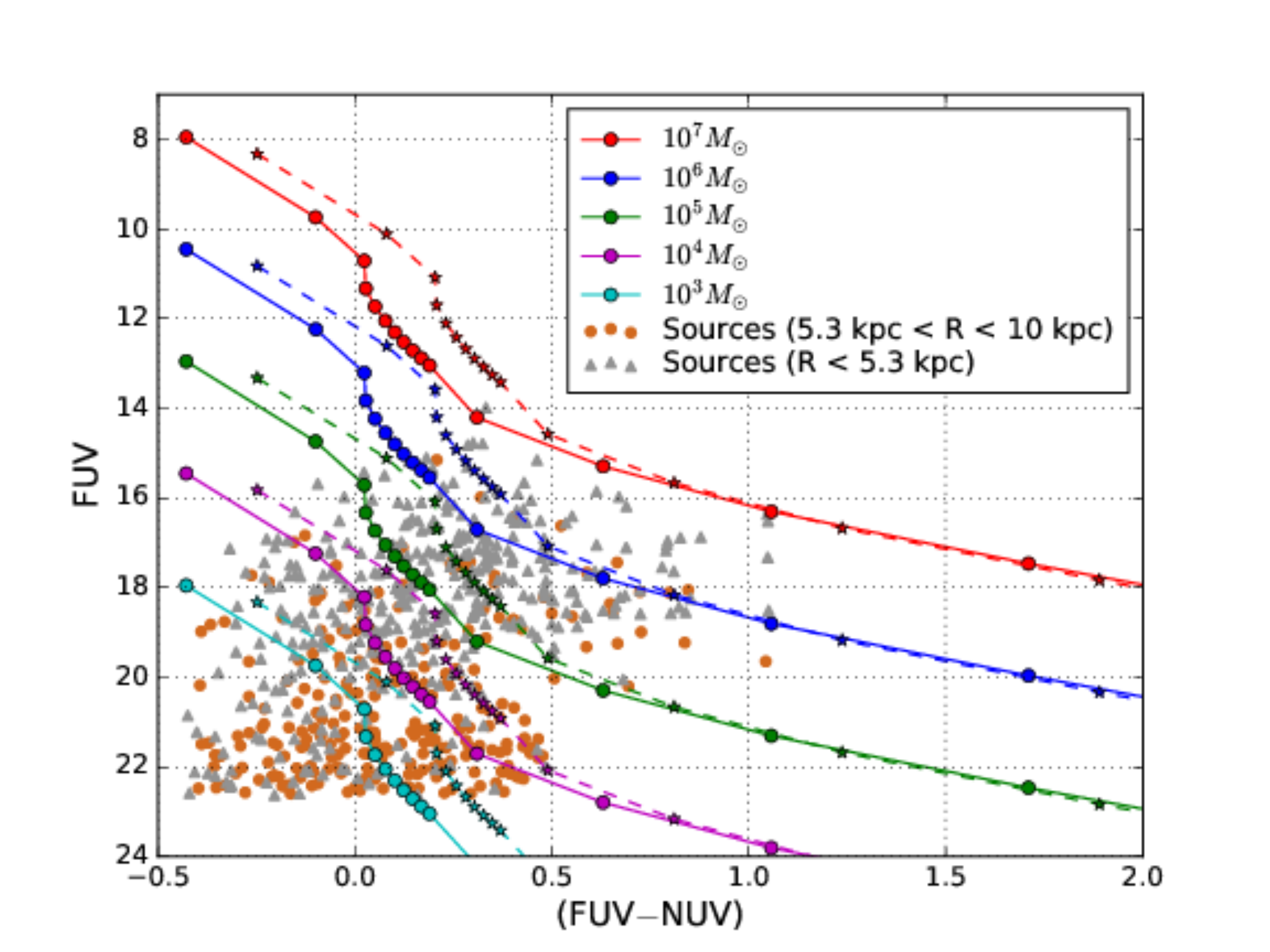} 
 \caption{The selected UV sources are shown on the simulated plot presented in Figure \ref{cmd_model}. The sources present in the outer disk between radii 5.3 kpc to 10 kpc are shown in brown circles whereas gray diamonds represent the sources in the inner disk within radius 5.3 kpc.}
 \label{cmd_galaxy}
 \end{center}
 \end{figure}

\subsection{Mass estimation of UV sources}
\begin{figure}
\begin{center}
\includegraphics[width=3.5in]{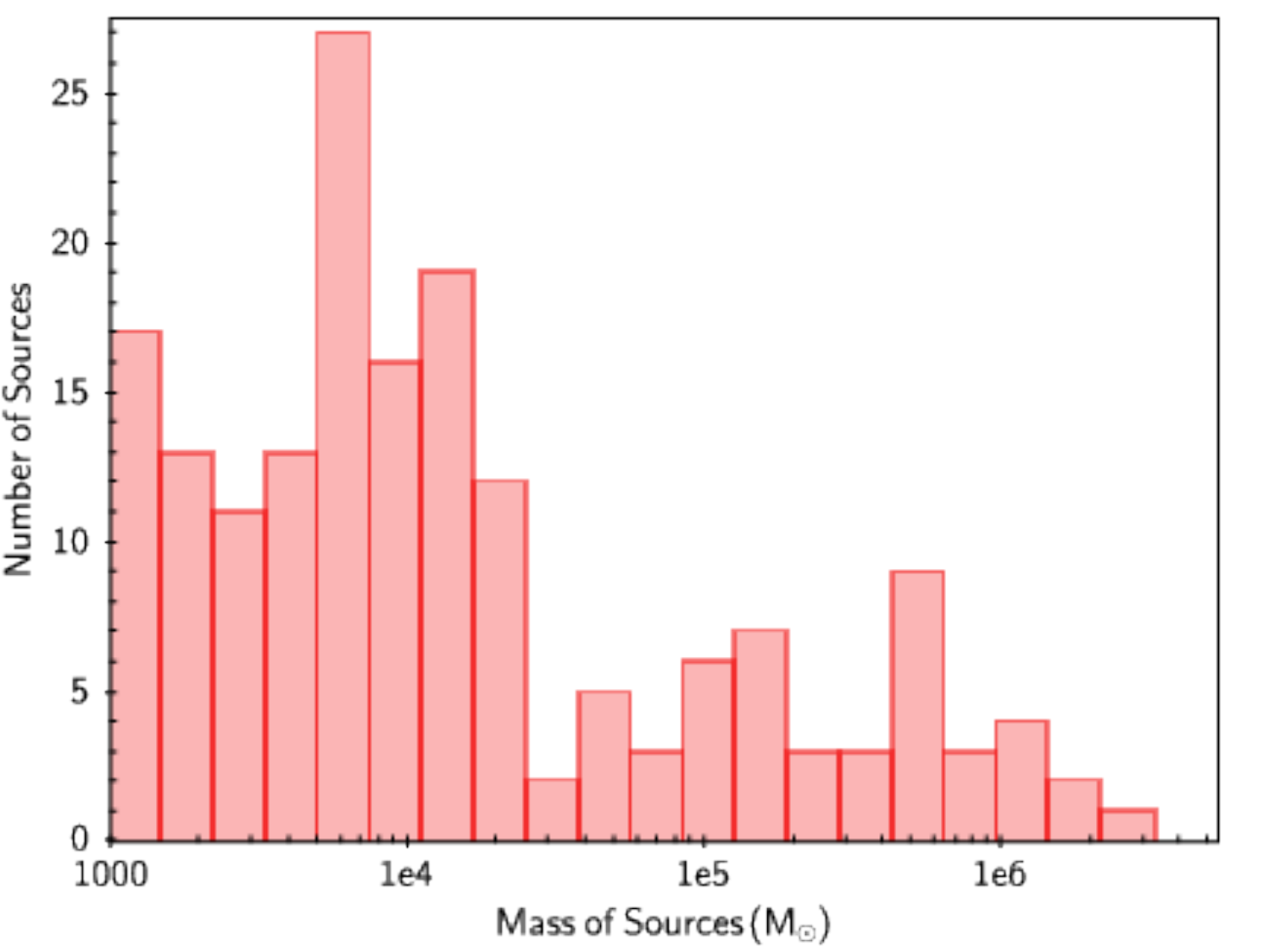} 
 \caption{Mass histogram of the detected UV sources present between radius 5.3 kpc and 10 kpc is shown. Sources with mass below $10^3 M_{\odot}$ are not displayed in the figure.}
 \label{mass_hist}
 \end{center}
 \end{figure}

The SSP models as shown in Figure \ref{cmd_model} suggest that for a given color 
(equivalently age), FUV magnitude 
changes with cluster mass. We plotted 742 selected sources on the simulated figure shown in Figure \ref{cmd_model} and displayed it in Figure \ref{cmd_galaxy}. This figure helps us to estimate mass for each selected source. We performed a linear interpolation between 
the two nearest model generated magnitude values to estimate the mass of a source corresponding to the observed extinction corrected 
magnitude. As our study aims to explore the outer disk only, we estimated mass for 261 sources present between radii 5.3 kpc to 10 kpc (1 $\sim$ 2 R$_{25}$)  of the galaxy. The histogram of the estimated masses of all these 261 sources is shown in Figure \ref{mass_hist}. The figure clearly shows that majority of the sources present in the outer disk of the galaxy have mass less than $10^5 M_{\odot}$. The sources with masses below $10^3 M_{\odot}$ are not shown in Figure \ref{mass_hist} as their measured value is not accurate due to the lower limit of model mass range (Figure \ref{cmd_model}). The sources identified in the inner disk are relatively more massive. There can be an artefact of assuming multiple sources as single one due to the crowding in the inner disk.\\

\subsection{Spatial mass distribution of UV sources}
\begin{figure}
\centering
\subfigure[]{\includegraphics[width = 3.2in]{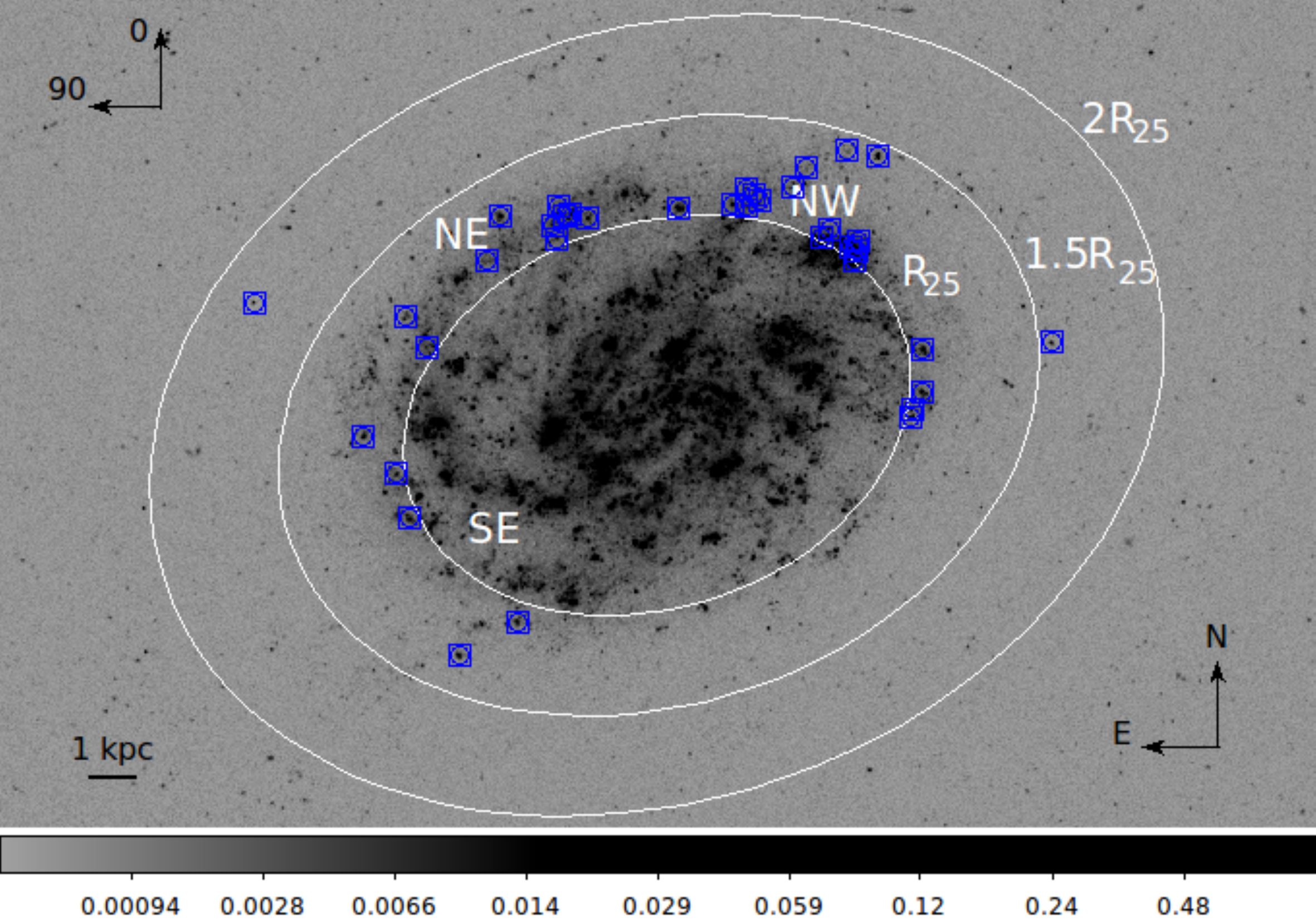}}
\subfigure[]{\includegraphics[width = 3.2in]{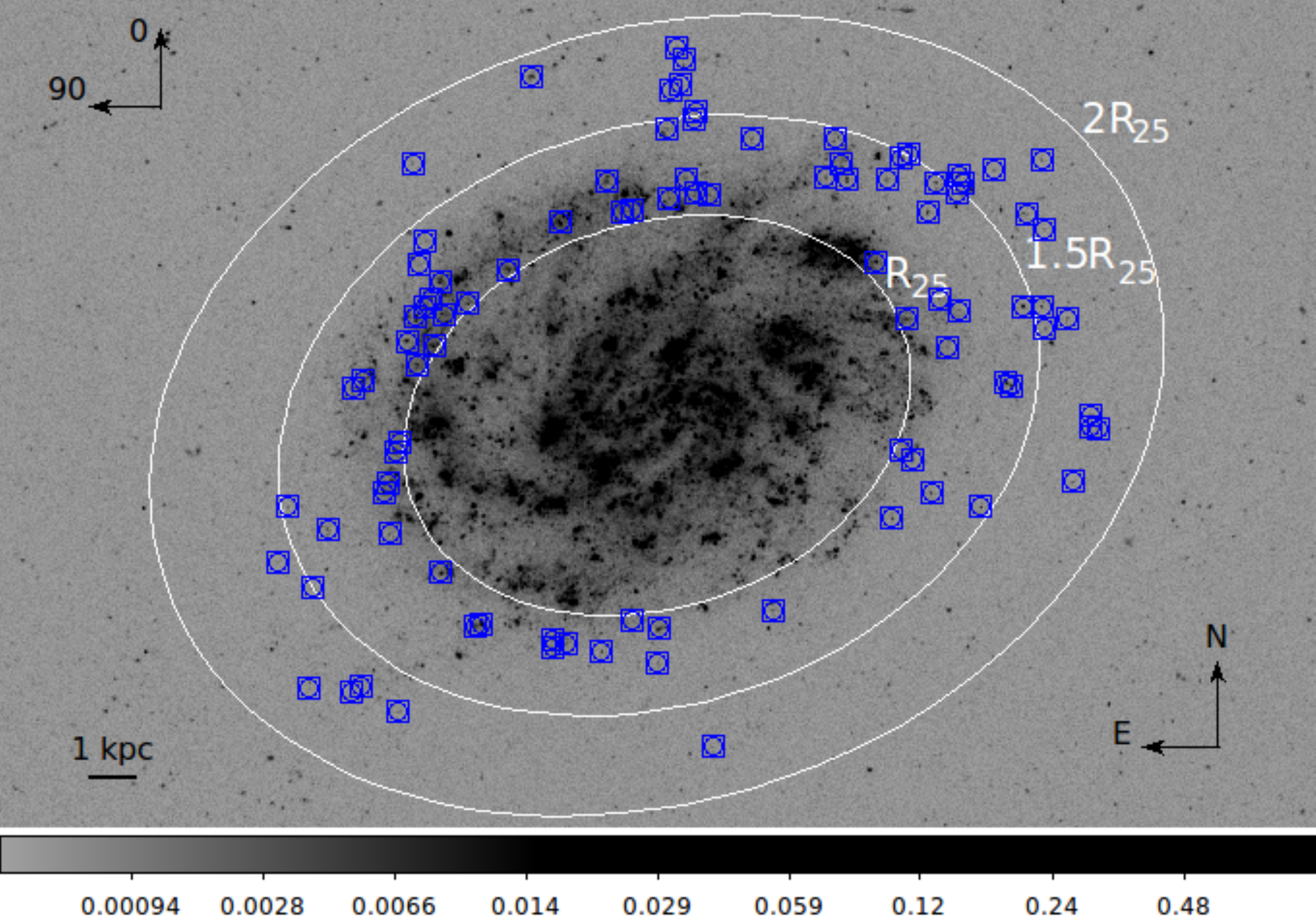}}
\subfigure[]{\includegraphics[width = 3.2in]{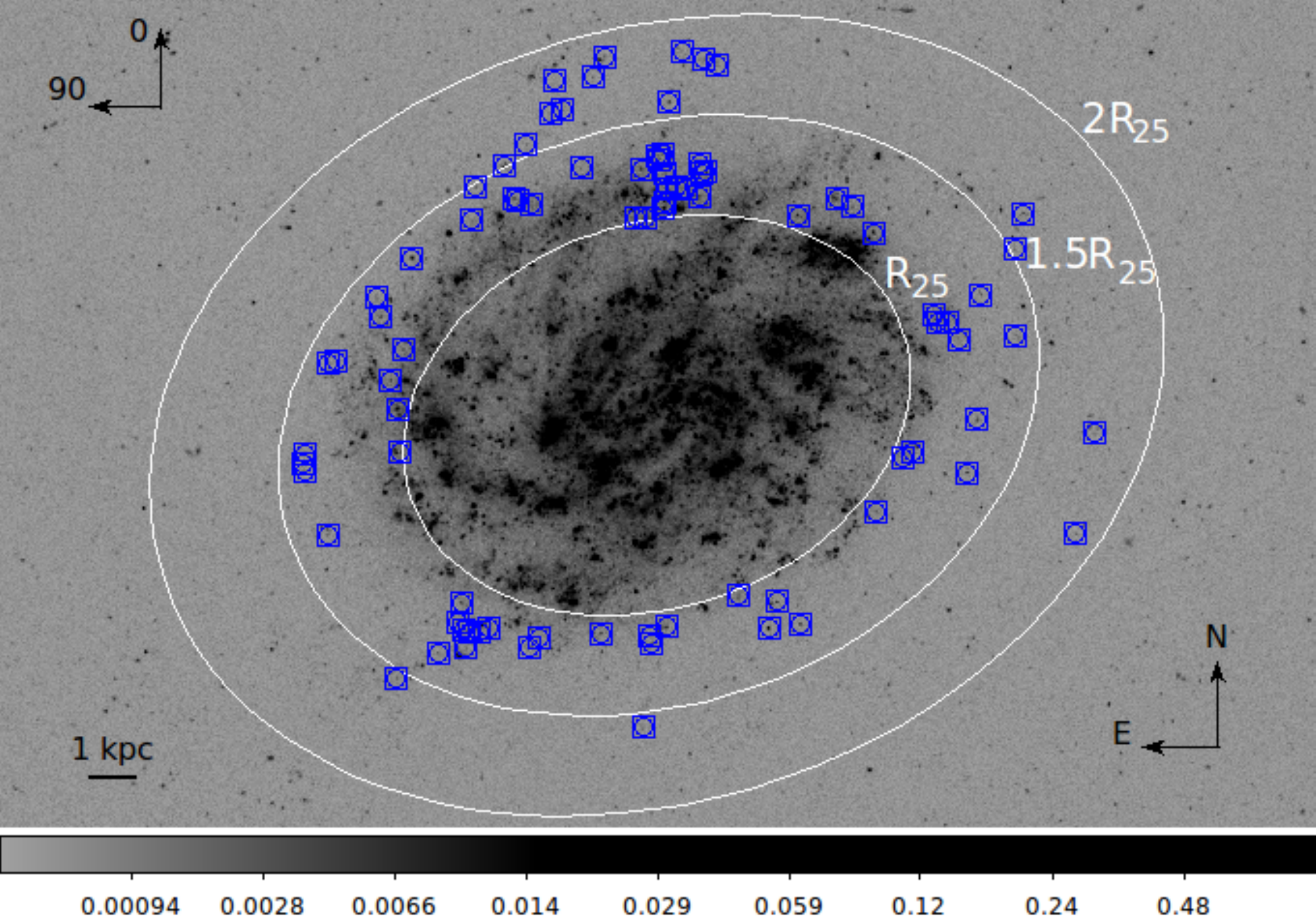}}
 \caption{The figures show the mass distribution of UV sources in the outer disk of the galaxy. Figure (a) shows the high mass (M$>10^5 M_{\odot}$) sources, Figure (b) shows intermediate mass ($10^3 M_{\odot}<$M$<10^5 M_{\odot}$) sources and low mass sources (M$<10^3 M_{\odot}$) are shown in Figure (c).}
 \label{mass_dist}
\end{figure}

In order to map the mass distribution of UV sources in the outer disk we have separated the selected sources in three groups, such as high mass (M $>10^5 M_{\odot}$), intermediate mass ($10^3 M_{\odot} < $M$ <10^5 M_{\odot}$) and low mass (M $< 10^3 M_{\odot}$) and over plotted them on GALEX FUV image in Figure \ref{mass_dist}. The massive sources are small in number and are mainly found near to the inner disk of the galaxy (Figure \ref{mass_dist}a). The extended arm in the north-west direction is found to have a few massive sources. We see this arm to be extended more in the outer disk in Figure \ref{mass_dist}b, where it is populated by intermediate mass sources. The inner part of north-eastern spiral arm is also found to have intermediate mass sources. The low mass sources, which are not shown in Figure \ref{mass_hist}, are mostly seen in the outer part of the north-eastern spiral arm (Figure \ref{mass_dist}c). We also noticed two clumps of low mass UV sources, one in the northern part and another in the south-eastern part of the galaxy. Figure \ref{mass_dist} conveys that the outer disk of NGC~300 has mostly formed low and intermediate mass sources ($\sim 86\%$ of the selected sample), in the last few hundred Myr. \\ 

\section{Discussion}
\label{s_discussion}
Understanding the nature of star formation in the outer part of disk galaxies is important ingredient to the evolution of the galaxy (Barnes et al. 2011). As the outer disks of galaxies have lower gas density, lower metallicity and lower dust content, altogether it offers an extreme environment to explore the characteristics of star formation (Zaritsky \& Christlein 2007; Barnes et al. 2011).  Detection of young star forming complex in the outer disk is also important to test the star formation threshold. The young and massive OB stars contribute a substantial amount of far-UV radiation which can trace the location of young star forming regions in a galaxy.  
The key aim of this study was to decipher the FUV disk properties of NGC~300 and to understand the age and mass distribution of UV sources in the outer disk of the galaxy.\\

The detection of several extended structures in the FUV  image strongly validates the presence of an outer disk in NGC~300. The galaxy is found to form substantially large number of objects during the last 25 Myr in its outer disk. The spiral arm present in the north-eastern direction is 
mainly populated by these young sources. Though we see a low level optical emission from this region 
of the galaxy, the detection of significant number of young sources in the UV image conveys that star
formation in the north-eastern spiral arm is a recent phenomenon. Rodr{\'i}guez et al. (2016) identified 1147
young stellar groups by studying six different regions of NGC~300 through HST observations. 
They reported an age range from 1 to 235 Myr for all the stars present in the blue bright groups.
Using another HST observation, Hillis et al. (2016) identified 576 stars with an age between 10 to 200 Myr in 
one of their chosen fields. We are also obtaining an age range of 1-300 Myr for the selected sources in our study.\\

The radial luminosity density profiles in different wavebands suggest that the disk of NGC~300 gradually becomes more steeper from shorter to longer wavelength. The scale-length estimated in FUV is found to be 2.3 times larger than that in infrared. This signifies that the older populations are more centrally concentrated in the galaxy disk whereas the younger ones show more extended distribution. This also confirms the presence of XUV disk in NGC~300. The radial profile of the H~I column density show a relatively flatter profile which suggests that star formation can trigger in the outer disk of the galaxy under favourable conditions.\\

Another important finding is the correlation between the H~I map and the identified young UV sources.
The north-eastern spiral arm, prominently seen in the UV image, is found to coincide with dense H~I gas.
A large fraction of the young sources are also identified along this arm. The
galaxy is believed to move in the south-east direction (Westmeier et al. 2011), which has
caused the compression of H~I gas in the south-east part of the disk.
The H~I contour map also shows a compression in the north-eastern direction, followed by dense H~I gas. The star
formation in the north-east region is likely to be due to this compression. We also detect a clump of low mass sources at the base of northern extended structure.\\

With the help of
starburst99 SSP model, we have analysed the GALEX photometric data of the
galaxy, NGC~300, and estimated the age and mass of identified
sources present in the outer disk. As these sources are likely to be star clusters, we derive the recent cluster
formation history of the galaxy.
As the formation of clusters is closely related to the formation of stars in general, the
cluster formation episodes are likely to
suggest star formation episodes as well.\\

The parameters estimated in this study are based on the SSP model. 
We have assumed instantaneous star formation and stellar IMF as Kroupa, which is appropriate to study star formation in nearby galaxies.
The errors of the estimated parameters (age and mass) are calculated from the photometric errors of both FUV and NUV magnitudes.
The range of photometric error for the (FUV$-$NUV) colour is 0.01 - 0.10. 
This error will reflect in the estimated
age of the sources. The error in age for most of the sources present in the outer disk is less than 20 Myr. Some of the young sources (age $<$ 150 Myr) show larger
error whereas sources older than 200 Myr have less error in estimated age. The photometric error in FUV magnitude will contribute to the
error in the estimation of mass. This error also has a range 0.01 - 0.10, which corresponds to 0.5 - 5 \% 
error in the estimated mass for all the sources.
The estimation of age has a dependency on reddening and metallicity. We have assumed moderate 
reddening and any decreased reddening will
make the ages younger. The metallicity for the entire disk is assumed to be Solar. If there is a 
significant gradient in the metallicity,
the age estimation will be affected by the dependency as can be seen in Figure \ref{GALEXmodel}. 
The estimation of
mass also has a dependence on the adopted extinction correction. As our study primarily explores the recent star forming regions, we have assumed a moderate
value of reddening and thus extinction. Any reduction in the assumed reddening will increase the FUV magnitude and the mass of a source and vice versa.\\

The outer disk of NGC~300 shows low level UV luminosity whereas optical studies indicate the presence of an unbroken stellar disk (Bland-Hawthorn et al. 2005). 
We detect several low mass sources in the outer disk of the galaxy. We found that the wispy extension of inner disk identified by Thilker et al. (2007) is due to these low mass young sources and slightly older intermediate mass sources. The low level star formation in the outer disk of NGC~300 may happen because of the less-availability of H~I. Mild perturbation due to the motion of the galaxy and ram pressure of the sculptor group medium could be causing the star formation in the outer disk. Therefore, this may be the process with which the outer disk is built in this galaxy.\\

The mass distribution of selected UV sources suggests that the outer disk has predominantly formed low and intermediate mass sources (M$<10^5 M_{\odot}$). Majority of the young sources (age $<$ 25 Myr) detected 
along the north-eastern spiral arm are found to have low mass which signifies a low level recent star formation in the galaxy. We considered Kroupa IMF to estimate mass of UV sources with the help of SSP model. Keeping other parameters as same (Table \ref{starburst999}), we changed the IMF value to 2.35 (classical value, Salpeter 1955) with a stellar
mass range from 0.1 - 120 $M_{\odot}$ and estimated the mass of the sources. The estimated masses were 1.3 - 1.8 times the mass estimated using 
kroupa IMF. The regions with intense 24 $\mu$m infrared emission in the galaxy spatially correlate with the massive star forming complexes detected in the inner disk of NGC~300. Faesi et al. (2014) studied 76 H~II regions in the disk of NGC~300 and estimated the stellar mass associated with 
each region. They found a range of mass from $10^3 M_{\odot}$ to $4\times10^4 M_{\odot}$.
The young stellar groups, identified by Rodr{\'i}guez et al. (2016), with substantially 
large number of stars are also likely to have total mass greater 
than $10^3 M_{\odot}$. In our study, we obtained a mass range from $10^3 M_{\odot}$ to $10^6 M_{\odot}$ (with some below $10^3 M_{\odot}$) for the sources present in the outer disk between 5.3 kpc and 10 kpc.\\

The nature of FUV luminosity density profile of NGC~300 points to the presence of an inner disk up to 5.5 kpc and an extended outer disk at least up to 12 kpc. The SFR of NGC~300 is found to be $\sim$ 0.46 $M_{\odot}/yr$. Considering the GALEX FUV data Verley et al. (2009) reported a SFR of 0.55 $M_{\odot}/yr$ for the galaxy M33. M33 is considered as the near optical twin of NGC~300 and they both have similar H~I mass. Therefore, the comparable values of SFR observed in both the galaxies signifies that they are undergoing a similar state of star formation. \\

\section{Summary}
\label{s_summary}
The main results of this study are summarized below:
\begin{enumerate}
\item Using GALEX UV data, we identified several extended structures in the outer part of the galaxy and confirmed the presence of an XUV disk in NGC~300.
\item The inner disk of the galaxy has a radius of 5.5 kpc whereas we detect an outer disk at least up to radius 12 kpc.
\item The disk scale-length, which found to increase gradually from longer to shorter wavelength, is estimated to be 2.66$\pm$0.20 kpc in NUV and  3.05$\pm$0.27 kpc in FUV.
\item We identified 261 candidate UV sources in the outer disk between radius 5.3 kpc to 10 kpc (1$\sim$2 R$_{25}$) and estimated their age and mass by applying SSP models.
\item We noticed a richness of younger (Age $<$ 25 Myr) as well as low and intermediate mass (M $<10^5 M_{\odot}$) sources in the outer disk of the galaxy.
\item The star formation in the north-eastern spiral arm of the galaxy is a recent phenomenon, consisting of low mass sources with age $<$ 25 Myr.
\item The distribution of UV sources identified in the outer disk correlates well with the features of H~I density map.
\item Presently, the galaxy is undergoing  a low level recent star formation in the outer disk ($\ge$ R$_{25}$), which may be due to its motion in the Sculptor group.
\item The SFR of NGC~300 ($\sim$0.46 $M_{\odot}/yr$) is found to be comparable with its near optical twin M33.
\end{enumerate}

\section*{Acknowledgements}
This study has primarily used ultra-violet data from GALEX observation. We thank both the GALEX and MAST team for providing science ready data products to the public. This research has also made use of the NASA/IPAC Extragalactic Database (NED), which is operated by the Jet Propulsion Laboratory, California Institute of Technology, under contract with the National Aeronautics and Space Administration. This research made use of Matplotlib (Hunter 2007), community-developed core Python package. Finally, we thank the referee for valuable suggestions.

\end{document}